\documentclass{aa}  

\usepackage{graphicx}
\usepackage{txfonts}
\usepackage{amsmath}	
\usepackage{amssymb}	
\usepackage{multirow}
\usepackage{float}
\usepackage{siunitx}
\usepackage{booktabs}
\usepackage{color,soul}
\usepackage[utf8]{inputenc}
\usepackage[flushleft]{threeparttable}  

\graphicspath{ {./../Astronomy_Astrophysic/} }
\setstcolor{red}

\newcommand{\Msun}{M$_{\hbox{$\odot$}}$}  
\newcommand{\Lsun}{L$_{\hbox{$\odot$}}$}  
\newcommand{\cmc}{cm$^{-3}$}              
\newcommand{\kms}{km\,s$^{-1}$}           
\newcommand{\mum}{$\mu$m} 
\newcommand{\Lco}{K\,\kms\,pc$^{2}$}
\newcommand{\Xco}{\hbox{M$_{\odot}$\,(K\,km\,s$^{-1}$\,pc$^2$)$^{-1}$}}

\begin{document}

   \title{CO(2-1) Survey at 9 parsec resolution in the SMC}


   \author{H. P. Salda\~no\inst{1},
          M. Rubio\inst{2},
          A. D. Bolatto\inst{3,}\inst{4,}\inst{5},
          C. Verdugo\inst{6},
          K. E. Jameson\inst{7},
          \and A. K. Leroy\inst{8,}\inst{9}
          }

   \institute{Observatorio Astron\'omico de C\'ordoba, Universidad Nacional de C\'ordoba, 5000 C\'ordoba, Argentina\\
              \email{hpablohugo@gmail.com}
         \and
             Departamento de Astronom\'ia, Universidad de Chile, Casilla 36-D, Santiago, Chile
        \and
             Department of Astronomy and Joint Space-Science Institute, University of Maryland, College Park, MD 20742, USA
        \and Visiting Scholar at the Flatiron Institute, Center for       Computational Astrophysics, NY 10010, USA
        \and Visiting Astronomer at the National Radio Astronomy          Observatory, VA 22903, USA
        \and
             Joint ALMA Observatory (JAO), Alonso de Córdova 3107, Vitacura, Santiago de Chile
        \and
             Space and Astronomy, CSIRO, 26 Dick Perry Avenue, Kensington, WA 6151, Australia
        \and
            Department of Astronomy, The Ohio State University, 140 West 18th Avenue, Columbus, Ohio 43210, USA
        \and
            Center for Cosmology and Astroparticle Physics, 191 West Woodruff Avenue, Columbus, OH 43210, USA
             }

   \date{Received September 15, 1996; accepted March 16, 1997}

 
  \abstract
   {The Small Magellanic Cloud (SMC) is the closest low-metallicity galaxy to the Milky Way where the dynamical state of molecular clouds can be analyzed. 
   }
   {Present a CO($2-1$) survey at $\sim$ 9 pc resolution obtained with the APEX telescope in an extensive region ($\sim$ 0.4 kpc$^{2}$) of the SMC and characterize the properties of the molecular clouds. Study the dynamical state and stability of these clouds uniformly and determine the link between molecular gas and star-forming regions.}
   {We use the CPROPS algorithm to identify the molecular clouds and estimate their main CO properties. We analyze the characteristic of the SMC clouds by studying the scaling relations between the radius, velocity dispersion, luminosity, and virial mass. We use the dust-based total gas mass of SMC clouds presented in the literature to analyze the stability of the molecular clouds. We use YSOs, HII region catalogs in the literature, and IR observations in public databases to inspect the star-forming regions in the SMC. We also analyze the cumulative mass distribution of the SMC molecular clouds.}
   {We identify 177 molecular clouds within the SMC, of which 124 clouds are fully resolved with signal-to-noise ratio $\gtrsim 5$. The scaling relationships show that the SMC clouds are (on average) less turbulent and less luminous than their inner Milky Way counterparts of similar size by a factor $\sim 2$ and $\sim 3$, respectively, while for a fixed linewidth, the SMC clouds are over-luminous by a factor $\sim 3.5$. \textcolor{black}{\bf The CO luminosity of the identified clouds is $(1.3 \pm 0.2)\times10^{5}$ \Lco\  (representing at least 70\% of the total CO luminosity of the region), and the corresponding gas mass from virial determination is M$_{\rm gas} = (1.5 \pm 0.5) \times10^6$ \Msun. Using the virial masses, we derive a CO-to-H$_2$ conversion factor for the \bf SMC CO clouds identified by CPROPS of $\alpha_{\text{CO}}^{\text{vir}} =$ \textbf{10.5}$\,\pm\,$\textbf{5} \Xco, measured at 9 pc resolution. Using literature gas masses from dust emission of a sub-sample of clouds for which we identify the corresponding CO emission, we  determine a dust-based conversion factor of $\alpha_{\text{CO}}^{\text{dust}} =$ \textbf{28}$\,\pm\,$\textbf{15} \Xco, obtained at 12 pc resolution. These conversion factors, determined by two alternative methods, are about $2.5$ and $6.5$ times larger than the canonical Galactic conversion factor. We study the stability of the molecular clouds where we have both a dust and a virial mass, and find that they appear to be in approximate gravitational virial equilibrium.} We find that the cumulative mass functions based on both the luminous mass and the virial mass are steeper than $dN/dM \propto M^{-2}$, suggesting that most of the molecular mass of the SMC is contained in low-mass clouds.
}
   {}


   \titlerunning{CO survey in the SMC}
   \authorrunning{H. P. Salda\~no, et al.}
   
   \maketitle
%

\section{Introduction}

The SMC, a nearby galaxy at $\sim 60$ kpc \citep[][]{Hilditch_2005MNRAS_357_304H}, is an ideal laboratory to study the link between the high-mass star formation and the interstellar medium (ISM) in low metallicity regions. Its ISM is characterized by a metallicity of $\sim$ 0.2 Z$_{\sun}$ \citep[12+$\log(\text{O/H}) \simeq$ 8.0,]
[]{Russell_1992ApJ_384_508R}, a high gas-to-dust ratios (GDR) of $\sim$\,\,2000 \citep{Roman_Duval_2017ApJ_841_72R} and a strong UV radiation field \citep[up to 10 times higher than that in the solar neighbourhood,][]{Vangioni_Flam_1980_AA_90_73V}. It contains a huge amount of atomic gas and an extremely low fraction of molecular components.
\textcolor{black}{\bf \citep[$\sim 4\%$ of the total mass of neutral gas (HI$+$He$+$metals) is molecular, see][]{Jameson_2016ApJ_825_12J,DiTeodoro_2019_MNRAS_483_392D}}.
All these characteristics of the ISM make the SMC an environment analogous to the early universe.

Measuring the amount of molecular hydrogen gas (H$_2$) in the SMC through observations of carbon monoxide (CO), the most common H$_2$ tracer, has been a real challenge in the last three decades due to the ISM characteristics. The first $^{12}$CO(1$-$0) survey in the SMC was done by \cite{Rubio_1991_ApJ_368_173R}, who mapped an area of 3$\times$2 degrees at 8.8 arcmin resolution (154 pc), finding large $^{12}$CO (hereafter CO) complexes of sizes $> 100$ pc from the South-West (SW) and the North-East (NE) of the Bar of the galaxy. Later, with better spatial resolutions of $\sim 40''$ ($\sim$ 12 pc at the SMC distance) and $22''$ (6.5 pc), an analysis of the CO($J=1-0$, $2-1$) and $^{13}$CO($J=1-0$, $2-1$) towards pointed and small areas in the SMC was performed by \cite{Israel_1993_A&A_276_25I},  \cite{Rubio_1993AA_271_1R,Rubio_1993AA_271_9R}, \cite{Rubio_1996AAS_118_263R} and \cite{Muller_2010ApJ_712_1248M}. These authors found CO clouds with sizes between $\sim10-40$ pc and peak emission between $\sim$0.3 to 7 K. A CO(1$-$0) survey of the entire SMC was performed by \cite{Mizuno_2001PASJ_53L_45M}, who mapped the emission in the SMC Bar and the SMC Wing (the transition to the Magellanic Bridge) with the NANTEN telescope, at a coarser spatial resolution of 2.6$'$ ($\sim 45$ pc). At much better resolution ($\sim 1-2$ pc), interferometric observations performed by the ALMA telescope identified several tiny CO clouds in different regions of the SMC \citep[e.g.,][]{Jameson_2018_ApJ_853_111J,Fukui_2020arXiv200513750F,Tokuda_2021ApJ_922_171T}.

These CO observations have unveiled very weak CO clouds with smaller beam-filling fractions than their Galactic counterparts, consistent with other low metallicity galaxies such as WLM \citep{Rubio_2015_Nature_525_218R} and NGC 6822 \citep{Schruba_2017ApJ_835_278S}. Different resolutions and methodologies employed in the study of the SMC CO emission, however, have made it difficult to obtain a homogeneous set of physical cloud properties. \textcolor{black}{\bf In coarse resolution observations \citep{Rubio_1991_ApJ_368_173R,Rubio_1993AA_271_9R,Mizuno_2001PASJ_53L_45M}, the SMC giant molecular clouds (GMCs) exhibit CO luminosity several times smaller than the inner Milky Way clouds for similar linewidth and virial mass, although they appear to be consistent with the Galactic size-linewidth relation. Later studies at higher resolution and sensitivity \citep{Bolatto_2008ApJ_686} using the CPROPS decomposition, however, show that the SMC clouds are under the Galactic size-linewidth relation and have CO luminosities similar to their virial mass. This suggests that measurements that reach the scale of the CO-dominated portion of GMCs yield conversion factors that are close to Galactic even in low metallicity environments such as the SMC.}

Environmental conditions also play a role in shaping the properties of clouds. The high pressures in galaxy centers may drive the equilibrium in the direction of larger velocity dispersions for a given size \citep{Oka_2001ApJ_562_348O,Colombo_2014ApJ_784_3C,Utomo_2015ApJ_803,Leroy_2015_ApJ_801_25L}. A similar result is found near the R136 stellar cluster in 30 Doradus in the LMC \citep{Kalari_2018_ApJ_852_71K}. Alternatively, environments of low pressure such as the outer Milky Way disk with small and transient clouds, display too large linewidths for their masses \citep{Heyer_2001ApJ_551_852H}. Evidence of the dependence of molecular parameters on the environment inside the SMC was investigated by \cite{Muller_2010ApJ_712_1248M}, finding that CO clouds are fainter and have narrower linewidths in the NE region than in the SW region. Note, however, that the dynamical properties of the molecular ISM depend on the spatial resolution \cite[e.g.,][]{Leroy_2016_ApJ_831_16L}. For this reason, homogeneous studies with common spatial resolution and uniform algorithms are important to obtain consistent results across a galaxy.

\begin{figure*}[ht!]
	\includegraphics{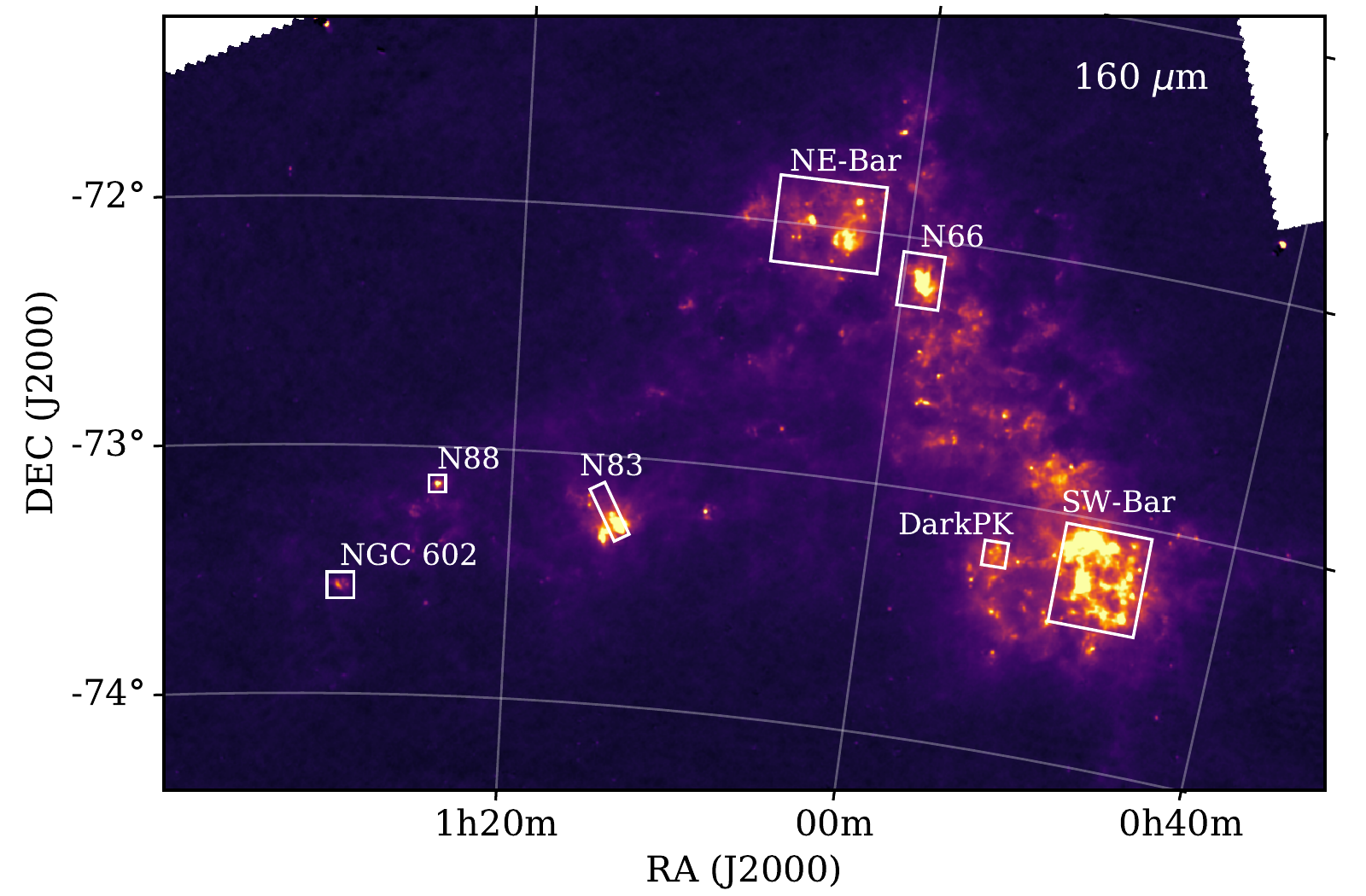}
    \caption{Image of the SMC at 160 \mum\ from the HERITAGE {\em Herschel} Space Observatory Key Project by \cite{Gordon_2014_ApJ_797_85G}, in which the Milky Way cirrus and background emission were subtracted. The white boxes outline the seven regions mapped with the APEX telescope.}
    \label{fig:smc_all_regions}
\end{figure*}

\subsection{The Survey}

Our goal is to characterize the CO clouds of the SMC at 9 pc resolution and examine their properties and dynamical states with the environment, for which we analyze seven different regions using CO(2$-$1) observations obtained with the Atacama Pathfinder Experiment (APEX) telescope. In Figure \ref{fig:smc_all_regions} we show the observed regions superimposed on the {\em Herschel} Space Observatory 160 \mum\ cold dust image \citep{Gordon_2014_ApJ_797_85G}. These regions are located towards the brightest dust emission sources and the highest H$_2$ column density of the galaxy \cite[see][]{Jameson_2016ApJ_825_12J}. The two largest regions observed are the SW-Bar and NE-Bar. We also mapped several smaller regions, N66 in the NE and DarkPK in the SW, and three regions located in the Wing of the SMC: N83, NGC 602, and N88. These maps cover regions with a large range of activity properties, the NE region is active but likely older and shows weaker clouds \citep{Muller_2010ApJ_712_1248M}, N66 is the brightest HII region of SMC with ongoing strong high-mass star formation \citep{Simon_2007ApJ_669_327S,Rubio_2018AA_615A_121R}, DarkPK is a quiescent region with a lot of molecular gas but very low activity dominated by "CO-dark" molecular gas \citep{Jameson_2018_ApJ_853_111J}, while N83, NGC 602 and N88, are isolated star-forming regions in the SMC-Wing. The high sensitivity and area coverage of these observations permit us to obtain a large uniform sample of CO clouds within the SMC.

We aim to analyze the GMC physical conditions using these homogeneous data with a consistent methodology. We will analyze the size-linewidth relation of the sample as well as other luminosity scaling relations ($L_{\text{CO}}-\sigma$ and $L_{\text{CO}}-R$) and the virial mass-Luminosity relation (M$_{\text{vir}}-L_{\text{CO}}$) which have been well defined for Galactic and extra-galactic GMCs \citep{Solomon_1987ApJ_319_730S,Bolatto_2008ApJ_686,Colombo_2014ApJ_784_3C,Leroy_2015_ApJ_801_25L}. These scaling relations will provide further constraints for cloud formation models \citep[e.g.,][]{Camacho_2016ApJ_833_113C}.

The Milky Way size-linewidth relationship is frequently used to study the dynamic and equilibrium turbulence conditions of molecular clouds. \cite{Solomon_1987ApJ_319_730S} find for inner Milky Way GMCs the following relation:

\begin{equation}
\centering
 \sigma_{\upsilon} = 0.72\,R^{\,0.5}\,\, \text{\kms,}
\label{eq:size-linewidth}
\end{equation}

\noindent
which indicates similar surface densities of clouds in gravitational virial equilibrium. This Galactic relation is also interpreted as evidence of equilibrium supersonic turbulence conditions in a highly compressible medium \citep{Bolatto_2013ARA&A_51}. Observations also suggest that molecular clouds in different galaxies generally follow power-law relationships, similar to those defined by \cite{Solomon_1987ApJ_319_730S} as:

\begin{equation}
\centering
 L_{\text{CO}} = 25\,R^{\,2.5}\,\, \text{\Lco,}
\label{eq:lco_radius_solomon}
\end{equation}

\noindent

\begin{equation}
\centering
 L_{\text{CO}} = 130\,\sigma_{\upsilon}^{\,5}\,\, \text{\Lco,}
\label{eq:lco-linewidth_solomon}
\end{equation}

\noindent
and 

\begin{equation}
\centering
 M_{\text{vir}} = 39\,L_{\text{CO}}^{\,0.8}\,\, \text{\Msun,}
\label{eq:Mvir_lco}
\end{equation}

\noindent
for the inner Milky Way clouds. In equation \ref{eq:lco_radius_solomon} and \ref{eq:lco-linewidth_solomon}, $R$ and $\sigma$ are expressed in pc and \kms, respectively, while equation \ref{eq:Mvir_lco} (the Galactic mass-luminosity relation) employs $L_{\text{CO}}$ in units of \Lco. 

In Section \ref{sec:APEX_observation}, we describe the APEX observations and data reduction, as well as the complementary data used in the paper. In Section \ref{sec:Methodoly}, we summarize the methodology to identify the molecular clouds and the main equations to determine the molecular parameters. In Section \ref{sec:main_Results}, we show the main results of the analysis of the scaling relations of the SMC clouds. We also examine the CO association with Young Stellar Objects (YSOs) and HII regions. In this section, we also present the analysis of the mass spectrum of the SMC through the cumulative mass distribution. In Section \ref{sec:discussion}, we discuss the dynamical states of the SMC molecular clouds, the CO-to-H$_2$ conversion factors of the galaxy, the observed features in the scaling relation of the SMC, and the comparison with those of other low-metallicity galaxies. Finally, in Section \ref{sec:summary}, we present a summary and conclusions.

\section{Observations}
\label{sec:APEX_observation}

\subsection{APEX Data}
\label{sec:APEX_data}

Seven regions of SMC, two in the North-East region: NE-Bar and N66; other two in the South-West region: SW-Bar and DarkPK; and other three in the Wing: N83, NGC 602, and N88  were observed with APEX, a 12 m diameter millimeter-wave telescope located in Llano de Chajnantor, Chile \citep{Gusten_2006_AA_454L_13G}.

The observations were done in the $^{12}$CO(2$-$1) molecular lines (230.538 GHz) using the Swedish Heterodyne Facility Instrument (SHeFI), APEX-1, under good weather conditions, in June 2014 for the SW region (C93.F-9711-2014) and during 2015 for the NE and Wing regions (C.095F-9705A-2015). The Half Power Beam Width (HPBW) of the APEX-1 instrument is $\sim$ 27.5$''$ at $^{12}$CO(2$-$1) frequency \citep{Vassilev_2008_AA_490_1157}. Typical system temperatures were between 120 to 200 K. All the observations were conducted using the total power on-the-fly mode, with a dump and grid spacing of 9$''$, and using a reference point free of CO(2$-$1) for background subtraction for each field, with a velocity resolution of 0.1 \kms, and sensitivity of 150 mK and 50 mK rms depending on the observed region.

The data reduction was performed according to the standard procedure of the CLASS software, Gildas\footnote{http://www.iram.fr/IRAMFR/GILDAS/}. The antenna temperature, $T_{A} $, was transformed to brightness temperature of the main beam ($ T_{mb} = T_{A} / \eta_{mb} $), using a main beam efficiency $\eta_{mb} = 0.7$  for APEX-1. In the final reduction, the data were obtained with a spatial resolution of HPBW $= 30''$ ($\sim 9$ pc) and smoothed to a common velocity resolution of $\Delta V = 0.25$ \kms\ to improve the signal-to-noise ratio ({\it rms}). Mainly linear polynomial was employed for baseline fitting except in some cases where a third-order polynomial was used. The {\it rms} of all regions are between $\sim 0.1 - 0.7$ K. Table \ref{obs_features} summarizes the corresponding observational parameters, including the equatorial coordinates and the field-of-view (FoV in col. 5) for each region.

The SW-Bar region was combined with additional APEX observations of three $180''\times\,600''$ smaller maps in the same area, namely N22 (RA$=$00:47:53.7, DEC$=-$73:17:54.8, PA$=$20), SWBarS (RA$=$00:45:23.4, DEC=$-$73:21:43.3, PA$=$10) and SWBarN (RA$=$00:48:17.4, DEC$=-$73:05:26.3, PA$=$145) sources,  performed in the same program C.93F-9711-2014. These maps, observed with the same velocity resolution $\Delta V = 0.1$ \kms\  as the SW-Bar big map but a higher sensitivity ($rms = 0.05$ K) were similarly reduced and smoothed to a velocity resolution of $\Delta V = 0.25$ \kms. The combination of these three maps with the SW-Bar full map improved the signal-to-noise ratio to about half of the $rms$ obtained from the SW map in the overlapping areas.

\begin{table*}
\footnotesize
\centering
\caption{Observational characteristics of the SMC regions}
\begin{threeparttable}
\begin{tabular}{lccccccc}
\hline 
\hline
SMC      &  regions  &     R.A.       &     DEC       &      FoV      &    P.A.   &  $V_{lsr}$   &  $T_{rms}$\tnote{a} \\
         &           & (hh:mm:ss.s) & (dd:mm:ss.s)  &  ($'\times'$) & (deg) &    (\kms)    &     (K)     \\
\hline
\multirow{2}{*}{SW}
& SW-Bar     &   00:46:54.3  &  $-$73:14:09.9  &  20.9$\,\times\,$24.6 & ~~~00  &   120.0 &    0.22   \\
& DarkPK    &    00:52:56.1  &  $-$73:12:31.7  &   5.5$\,\times\,$5.8  & ~~~00  &   150.0 &    0.10   \\
\hline
\multirow{2}{*}{NE}
& NE-Bar     &   01:04:13.6  &  $-$71:59:23.3  &  24.8$\,\times\,$19.7 & ~~~00  &  167.0 &    0.24   \\
& N66       &    00:59:06.1  &  $-$72:10:05.7  &  \,\,\,9.0$\,\times\,$11.9 & ~~~00   &   160.0 &    0.30   \\
\hline
\multirow{3}{*}{Wing}
& N83       &   01:14:30.0   &  $-$73:13:26.8  &  \,\,\,4.1$\,\times\,$13.8 & $+$30 & 161.7 &    0.10   \\
& NGC 602  &   01:29:26.8  &  $-$73:33:51.4  &   5.8$\,\times\,$5.8   & ~~~00  &   166.5 &    0.30   \\
& N88      &   01:24:08.6  &  $-$73:08:53.6  &   3.0$\,\times\,$3.0   & ~~~00  & 148.0 &    0.70   \\
\hline
\end{tabular}
\begin{tablenotes}
\item[a] The main brightness temperature $T_{rms}$ is calculated in  $0.25$~km\,s$^{-1}$ spectral resolution.   
\end{tablenotes}
\end{threeparttable}
\label{obs_features}
\end{table*}

\subsection{Complementary Data}
\label{sec:IR_data}

We compare our CO observation with indicators of recent star formation in the SMC. For this comparison we use the Infrared Array Camera (IRAC) 8 \mum\ map from the Spitzer Survey ``Surveying the Agents of Galaxy Evolution" \citep[SAGE;][]{Gordon_2011_AJ_142_102G}. The 8 \mum\ emission is dominated by PAH bands excited by high UV radiation fields coming from massive stars creating photo-dissociated regions (PDRs). We also use an H$\alpha$ image from the ``Magellanic Cloud Emission-Line Survey"  \citep[MCELS,][]{Winkler_2015ASPC_491_343W} to compare the CO emission with ionized interstellar structures. The far-infrared (FIR) and the millimeter (mm) emissions are also used as important indicators of star formation due to their large fractions of dust-absorbed starlight. We use the FIR map at 160 \mum\ from the HERITAGE Herschel Key Project \citep[][]{Gordon_2014_ApJ_797_85G}, where the Milky Way cirrus and the background have been subtracted by these authors. We also use the AzTEC  1.1 mm dust continuum survey of the SMC carried out by \cite{Takekoshi_2017ApJ_835_55T} at a Gaussian FWHM of 40$''$.

\section{Methodology}
\label{sec:Methodoly}

In all regions, we apply the CPROPS\footnote{https://people.ok.ubc.ca/erosolo/cprops/} package developed by \cite{Rosolowsky_2006PASP_118_590} to identify and decompose the CO distribution into molecular clouds. In Appendix \ref{app:cprops_aplication}, the identification and decomposition procedure applied to our data cubes are explained in more detail. Once the clouds are identified, CPROPS determine their main properties, such as central velocities, sizes, velocity dispersion, and fluxes through the moment method. This method uses the distribution of the emission in a cloud within a position-position-velocity data cube and does not assume a previous functional shape of the cloud. CPROPS corrects the quoted properties by sensitivity and convolution biases.

To determine the size of the clouds, we use the \cite{Solomon_1987ApJ_319_730S} definition: $R \simeq 1.91\,\sigma_{r}$, where $\sigma_{r} = \sqrt{\sigma_{\text{maj}}\,\sigma_{\text{min}}}$ and $\sigma_{\text{maj}}$ and $\sigma_{\text{min}}$ are the rms sizes of the intensity distribution along the two spatial dimensions \citep[equation 9 from][]{Rosolowsky_2006PASP_118_590}. For those clouds that are more elongated in one axis than the telescope beam (extrapolated $\sigma_{\text{min}} < \sigma_{\text{beam}}$) and in consequence not well resolved, CPROPS cannot estimate their sizes. In such cases, we assume that the extrapolated $\sigma_{\text{min}} = \sigma_{\text{beam}}$ as an upper limit in this axis, and the new rms size is given by $\sigma_{r} = \sqrt{\sigma_{\text{maj}}\,\sigma_{\text{beam}}}$, which is then deconvolved in the standard way:

\begin{equation}
\label{eq:upper_deconv}
 \sigma_{r\text{,upp}} = \sqrt{\sigma^{2}_{r} - \sigma^{2}_{\text{beam}}}.
\end{equation}

\noindent
This upper limit rms size is also multiplied by 1.91 to estimate the upper limit radius ($R_{\text{upp}}$) of clouds. The velocity dispersion ($\sigma_{\upsilon}$) of each cloud, is calculated through equation 10 from \cite{Rosolowsky_2006PASP_118_590} in \kms\ unit. 

The luminosity of the molecular gas is given by:

\begin{equation}
\label{eq:luminosity}
\begin{aligned}
 L_{\text{CO}(1-0)}[\text{K\,km\,s}^{-1}\,\text{pc}^{2}] = \,   
  & F_{\text{CO}(1-0)}(0\,\text{K})[\text{K\,km\,s}^{-1}\,\text{arcsec}^{2}] 
   (d[\text{pc}])^{2} \\
  &  \times \left(\frac{\pi}{180\,\times\,3600}\right)^{2}
\end{aligned}
\end{equation}

\noindent
where $F_{\text{CO}(1-0)}(0\,\text{K})$ is the flux of the cloud extrapolated to $T_{rms} = 0$ K, and $d$ is the distance of the galaxy in parsec. In the following, we consider that the CO($2-1$)/CO($1-0$) integrated line brightness ratio is, on average, equal to $1.0$ in the Magellanic Clouds \citep{Rubio_2000AA_359_1139R,Israel_2003AA_406_817I,Bolatto_2003ApJ_595_167B,Herrera_2013AA_554A_91H,Tokuda_2021ApJ_922_171T}, with variation between $\sim 10-30$\%. Thus, with this measured ratios, $L_{\text{CO}(2-1)} \simeq L_{\text{CO}(1-0)}$. 

Another important parameter is the virial mass of the clouds. Assuming that clouds can be approximated by a self-gravitating sphere with a density profile $\rho \propto r^{-1}$ and discarding external force supports, such as magnetic fields and/or external pressure, the virial mass is given by the formula \citep{Solomon_1987ApJ_319_730S}: 

\begin{equation}
    M_{\text{vir}} = 1040\,\sigma_{\upsilon}^{2}\,R\,\,[M_{\odot}],
    \label{eq:virial_mass}
\end{equation}

\noindent
where $\sigma_{\upsilon}$ and $R$ are the corrected velocity dispersion in \kms\ and radius in parsec, respectively. We also calculate the luminous mass of the clouds by using the $L_{\text{CO}}$ and the Galactic CO-to-H$_2$ conversion factor $\alpha_{\text{CO}} =  4.36$ \Xco \citep[which includes the He mass contribution,][]{Bolatto_2013ARA&A_51} as follows:

\begin{equation}
    M_{\text{lum}} = L_{\text{CO}}\,\alpha_{\text{CO}}.
    \label{eq:luminous_mass}
\end{equation}

To determine the uncertainties of the parameters, the bootstrapping technique was used through CPROPS adding the BOOTSTRAP keywords. For those clouds not well resolved we use the error propagation to estimate the error in the upper limit radius and virial mass. 

\section{CO Analysis and Results}
\label{sec:main_Results}

\begin{figure*}
    \centering
	\includegraphics[width=\linewidth]{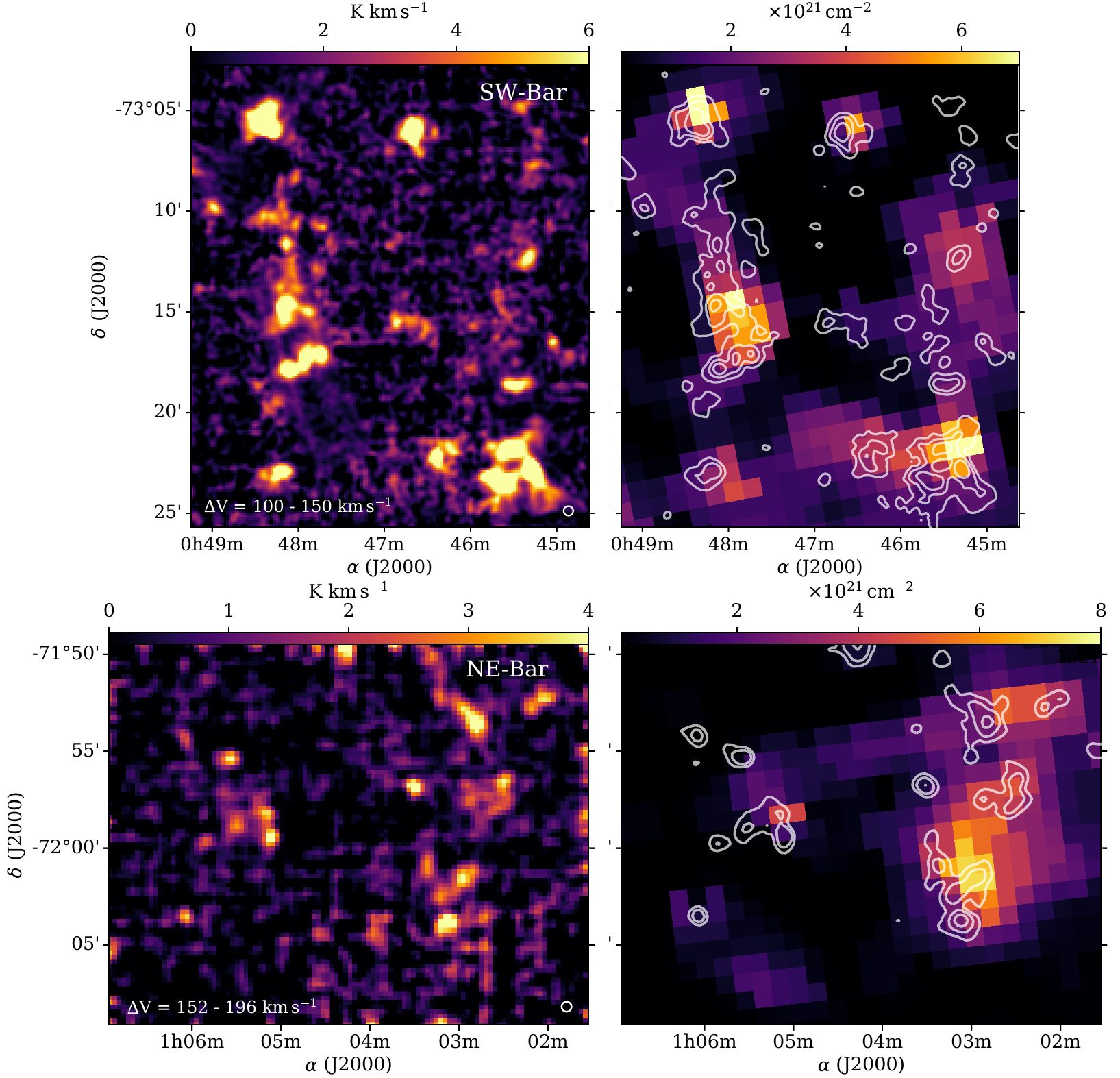}
    \caption{Integrated CO(2$-$1) emission toward the SW-Bar (top) and NE-Bar (bottom) regions. Only for this figure both CO maps were smoothed with a Gaussian kernel of a radius of 2 pixels to improve the signal-to-noise ratio. The beam size (30$''$ or $\sim$ 9 pc at the SMC distance) and the integration range ($\Delta$V) are indicated at the bottom of the left panels. In the right panels, the integrated CO contours (determined with the Moment Masked Method, see Section \ref{sec:main_Results}) are overlapped to the dust-based total column density (N$_{\text{H}_{2}}$) from \cite{Jameson_2016ApJ_825_12J}. The CO contours in the SW and NE regions correspond to emissions at 1.5, 4.5, 10.0 K\,\kms, and 1.0, 2.0, 4.5 K\,\kms, respectively. }
    \label{fig:CO_SW_NE}
\end{figure*}

\begin{figure*}
    \centering
	\includegraphics[width=\linewidth]{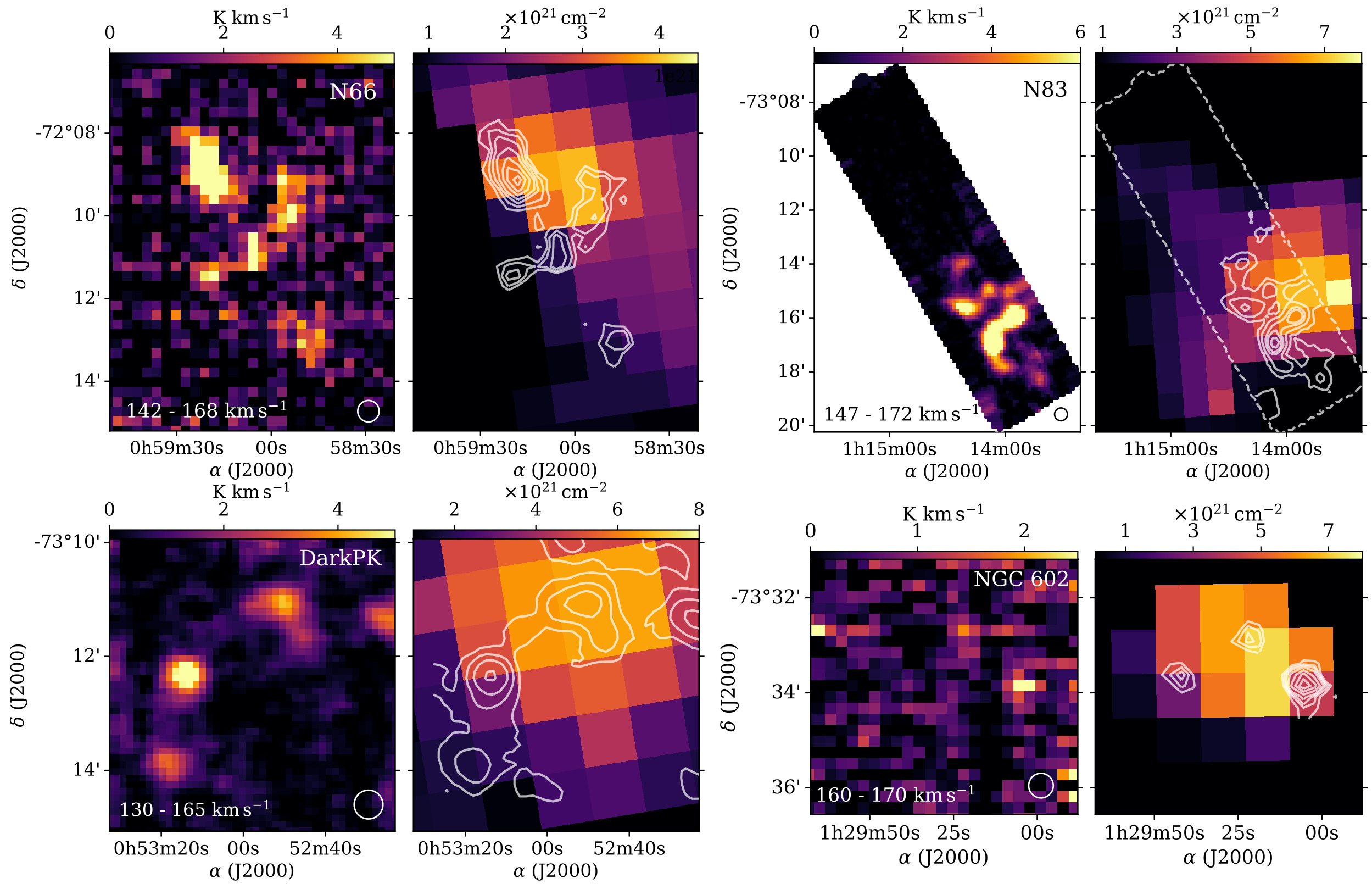}
    \caption{Integrated CO(2$-$1) emission toward N66, N83,  DarkPK, and NGC 602. The beam width of 30$''$ ($\sim$ 9 pc at the SMC distance) and the integrated velocity range are shown at the bottom of the left panels of each region. The CO contours in the right panel of N66 are at 1.6,  2.6,  4.0,   5.7,  8.6, 11.4, and 14.3 K\,\kms. While in N83, the contours are at 1, 3, 7, 10, and 15 K\,\kms. In DarkPK, the CO contours take values of 0.5, 1.5, 3.0, and 6.0 K\,\kms, and in NGC 602, 0.4, 0.5, 0.6, 0.9, 1.3, 1.7, and 2.1 K\,\kms. All these contours were determined with the Moment Mask Method (see Section \ref{sec:main_Results}).}
    \label{fig:CO_smallregions}
\end{figure*}

The left panels of Figure \ref{fig:CO_SW_NE} and \ref{fig:CO_smallregions} show the CO emission in the observed regions of the SMC. These CO maps are obtained by integrating within velocity ranges which include all molecular clouds detected in each region. For SW-Bar and NE-Bar, the velocity ranges are $100-150$\,\kms\ and $152-196$\,\kms, respectively. While for N66 and N83, the velocity ranges are $142-168$\,\kms\ and $147-172$\,\kms, and finally, DarkPk and NGC 602 are integrated between $130-165$\,\kms, and $160-170$\,\kms, respectively. All velocities are in $V_{lsr}$.  Our observations also include the N88 region with a single CO cloud detected between $145-150$ \kms. N88 is shown in Appendix \ref{app:CO_DUST_emission}.

To better display the CO emission in Figures 2 and 3,  we use the Moment Masking Method optimized for CO spectral line surveys \cite[for a detailed description of the method, see][]{Dame_2011arXiv1101_1499D}. With this method, small CO clouds with narrow velocity linewidth (few spectral channels) are recovered by integration over large velocity ranges. The moment-masked CO emissions are shown as contours in Figures \ref{fig:CO_SW_NE} and \ref{fig:CO_smallregions}, superimposed on the dust-based H$_2$ column density (N$_{\text{H}_{2}}$) map at $\sim$ 10 parsec resolution from \cite{Jameson_2016ApJ_825_12J}, except for N88 where no H$_2$ column density map has been derived.

The CO distribution in the SW-Bar and NE-Bar region, both obtained within a similar area of $\sim$ 500 arcmin$^2$ ($\sim 0.15$ kpc$^{2}$) and $rms \sim 0.2$ K,  show significant differences, as previously noted by  \cite{Muller_2010ApJ_712_1248M}. The CO emission towards the SW-Bar region is stronger and more extended than that of the NE-Bar region. In general, Figures \ref{fig:CO_SW_NE} and \ref{fig:CO_smallregions} show  CO emission towards high H$_2$ column densities. Interestingly, additional small isolated molecular clouds are detected for the first time in this work which is not present in the $N_{\text{H}_2}$ map. 

\subsection{Molecular Cloud Parameters}
\label{sec:molecular_clouds_parameters}

In this section, we analyze the properties of the molecular clouds identified by CPROPS \citep{Rosolowsky_2006PASP_118_590}. The physical parameters of the CO clouds are listed in Table \ref{tab:cprops_parameters_SMC}. In col. 2 and 3, the Right Ascension and Declination in $J(2000)$ are indicated. In col. 4 to 10, we list the radius ($R$), the {\it local-standard of rest} velocity ($V_{lsr}$), the velocity dispersion ($\sigma_{\upsilon}$), the peak temperature ($T_{\text{peak}}$), the integrated CO($2-1$) intensity ($I_{\text{CO}}$), the CO($2-1$) luminosity ($L_{\text{CO}}$), and the virial mass ($M_{\text{vir}}$). All these parameters obtained using CPROPS are corrected by sensitivity and resolution bias. To avoid redundancy in the analysis of these parameters, and for simplicity, we refer as South-West (SW) clouds to those that belong to the SW-Bar and DarkPK, as North-East (NE) clouds to those of the NE-Bar and N66, and as Wing (WG) clouds to those belonging to N83, NGC 602 and N88. In this analysis, we identify 177 CO clouds in the SMC, with 102 clouds in the SW, 47 clouds in the NE, and 28 clouds in the WG. \textcolor{black}{\bf The total luminosity of the 177 CO clouds is $1.3\times10^{5}$ \Lco.  To estimate the total CO emission irrespective of whether it is in CPROPS identified clouds we produce an integrated intensity map including all beams with a  signal-to-noise ratio (SNR)$>3$. The resulting total luminosity in the region is  $1.9\times10^{5}$ \Lco, likely an upper limit since the presence of noise will tend to bias the estimate high (although probably not substantially). 
Comparing these two numbers, we estimate that at most $\sim30$\% of  the CO emission is not assigned to CO clouds.}

Out of the total sample, 124 clouds (70\%)  are well-resolved with non-deconvolved radii of $R \gtrsim 5.0$ pc and SNR higher than 5.0. A radius of 5.0 pc corresponds to a deconvolved radius of $R \simeq 2.2$ pc, which will be considered as our size resolution limit. The relative error in $R$ of these well-resolved clouds \textcolor{black}{are typically between $10-30$\% with only a few clouds exceeding these values.} Clouds with smaller radii than $2.2$ pc have high uncertainties ($\sim 100\%$ or higher), \textcolor{black}{which strongly affect  their virial mass determination, and  therefore, their $R$ and $M_{\text{vir}}$ are considered as undefined values} in Table \ref{tab:cprops_parameters_SMC}. Another 28 clouds have SNRs between 3 and 5 and thus the accuracy of the parameter determination deteriorates producing large underestimations in their radius and linewidth. In the following sections, we will use in the analysis only the \textcolor{black}{124} well-resolved clouds (deconvolved $R \gtrsim 2.2$ pc and SNR > 5).

\begin{table*}
\centering
\caption{Physical parameters of molecular clouds in the SMC}
\begin{threeparttable}
\begin{tabular}{lcc
                S[table-format=1.1]@{\,\( \pm \)\,}
                S[table-format=1.1]
                S[table-format=3.1]@{\,\( \pm \)\,}
                S[table-format=1.1]                
                S[table-format=1.1]@{\,\( \pm \)\,}
                S[table-format=1.1]
                S[table-format=1.1]@{\,\( \pm \)\,}
               S[table-format=1.1]
                S[table-format=3.1]@{\,\(\pm\)\,}
                S[table-format=2.1]
                S[table-format=2.1]@{\,\(\pm\)\,}
                S[table-format=1.1]
                S[table-format=2.1]@{\,\( \pm \)\,}
                S[table-format=1.1]}
\hline
\hline
  ID  &    R.A.    &    Decl.   & 
  \multicolumn{2}{c}{$R$}                  &
  \multicolumn{2}{c}{$V_{\text{lsr}}$}     &
  \multicolumn{2}{c}{$\sigma_{\upsilon}$}  &
  \multicolumn{2}{c}{$T_{\text{peak}}$}    &
  \multicolumn{2}{c}{$I_{\text{CO}(2-1)}$}      &
  \multicolumn{2}{c}{$L_{\text{CO}(2-1)}$}      &
  \multicolumn{2}{c}{$M_{\text{vir}}$}     \\
      & (hh:mm:ss.s) & (dd:mm:ss) & 
      \multicolumn{2}{c}{(pc)}      &
      \multicolumn{2}{c}{(\kms)}    &
      \multicolumn{2}{c}{(\kms)}    & 
      \multicolumn{2}{c}{(K)}       &
      \multicolumn{2}{c}{(K\,\kms)} &
      \multicolumn{2}{c}{(caption)} &
      \multicolumn{2}{c}{(10$^{3}$\Msun)} \\
\hline
  1 & 00:44:57.6 & -73:10:07 &  2.9 & 1.1 & 132.3 & 1.1 & 1.0 & 0.3 &  1.5 &  0.2 &   66.7 & 14.7 &   2.8 & 0.6 &   3.0 &  2.0 \\
  2 & 00:45:00.4 & -73:16:45 & 12.5 & 1.3 & 108.6 & 1.4 & 1.2 & 0.1 &  2.0 &  0.2 &  289.3 & 20.4 &  12.2 & 0.9 &  18.7 &  4.3 \\
  3 & 00:45:02.5 & -73:24:06 & 10.8 & 0.9 & 126.4 & 1.4 & 1.2 & 0.1 &  1.8 &  0.2 &  252.5 & 17.9 &  10.7 & 0.8 &  16.2 &  3.4 \\
  4 & 00:45:08.7 & -73:20:37 &  \multicolumn{2}{c}{...} & 125.8 & 1.2 & 1.1 & 0.3 &  0.9 &  0.2 &   45.5 & 12.6 &   1.9 & 0.5 &   \multicolumn{2}{c}{...} \\
  5 & 00:45:10.5 & -73:24:39 &  6.8 & 1.3 & 128.1 & 1.2 & 1.0 & 0.1 &  1.5 &  0.2 &  118.6 & 10.7 &   5.0 & 0.5 &   7.1 &  2.4 \\
  6 & 00:45:11.8 & -73:21:21 &  4.3 & 0.9 & 122.8 & 1.0 & 0.8 & 0.3 &  1.1 &  0.2 &   74.2 & 30.8 &   3.1 & 1.3 &   2.9 &  2.7 \\
  7 & 00:45:14.9 & -73:06:13 &  6.4 & 0.8 & 102.7 & 1.2 & 1.0 & 0.2 &  1.7 &  0.3 &  119.6 & 24.1 &   5.1 & 1.0 &   6.7 &  3.2 \\
  8 & 00:45:16.1 & -73:20:27 &  2.5 & 1.2 & 120.9 & 1.7 & 1.5 & 0.4 &  0.9 &  0.2 &   45.1 & 14.1 &   1.9 & 0.6 &   6.1 &  3.9 \\
  9 & 00:45:16.4 & -73:23:07 & 13.0 & 0.6 & 126.9 & 2.9 & 2.5 & 0.1 &  3.8 &  0.2 & 1358.6 & 47.7 &  57.5 & 2.0 &  84.5 &  7.6 \\
 10 & 00:45:18.1 & -73:08:35 &  2.6 & 1.2 & 110.7 & 1.0 & 0.9 & 0.2 &  1.5 &  0.3 &   52.6 & 14.3 &   2.2 & 0.6 &   2.2 &  1.5 \\
 11 & 00:45:18.3 & -73:07:44 &  \multicolumn{2}{c}{...} & 108.2 & 0.8 & 0.7 & 0.5 &  1.1 &  0.3 &   32.2 & 16.9 &   1.4 & 0.7 &   \multicolumn{2}{c}{...} \\
 12 & 00:45:18.6 & -73:07:45 &  5.4 & 1.8 & 103.5 & 1.9 & 1.6 & 0.3 &  1.8 &  0.3 &  134.2 & 19.0 &   5.7 & 0.8 &  14.4 &  7.7 \\
 13 & 00:45:20.4 & -73:21:06 &  3.3 & 1.0 & 116.0 & 1.3 & 1.1 & 0.1 &  2.1 &  0.2 &  154.8 & 10.5 &   6.5 & 0.4 &   4.2 &  1.7 \\
 14 & 00:45:21.2 & -73:12:20 &  6.2 & 0.8 & 123.1 & 1.3 & 1.1 & 0.1 &  3.5 &  0.3 &  339.7 & 19.8 &  14.4 & 0.8 &   7.8 &  2.0 \\
 15 & 00:45:21.3 & -73:18:27 &  \multicolumn{2}{c}{...} & 116.3 & 1.4 & 1.2 & 0.2 &  1.5 &  0.2 &   85.9 &  9.1 &   3.6 & 0.4 &   \multicolumn{2}{c}{...} \\
 ... & ... & ... &  \multicolumn{2}{c}{...} & \multicolumn{2}{c}{...} &  \multicolumn{2}{c}{...} &  \multicolumn{2}{c}{...} &  \multicolumn{2}{c}{...} &  \multicolumn{2}{c}{...} &  \multicolumn{2}{c}{...} \\
 ... & ... & ... &  \multicolumn{2}{c}{...} & \multicolumn{2}{c}{...} &  \multicolumn{2}{c}{...} &  \multicolumn{2}{c}{...} &  \multicolumn{2}{c}{...} &  \multicolumn{2}{c}{...} &  \multicolumn{2}{c}{...} \\
176 & 01:29:20.7 & -73:32:52 &  3.8 & 3.2 & 162.0 & 0.4 & 0.4 & 0.2 &  0.9 &  0.2 &    9.8 &  2.6 &   1.7 & 0.4 &   0.5 &  0.8 \\
177 & 01:29:42.4 & -73:33:46 &  2.8 & 1.3 & 166.8 & 0.5 & 0.5 & 0.2 &  1.1 &  0.2 &    5.3 &  1.4 &   0.9 & 0.2 &   0.7 &  0.7 \\
\hline
\hline
\end{tabular}
\begin{tablenotes}
   \item {Note:} $L_{\text{CO}}$ is in unit of $10^{2}$\,K\,\kms\,pc$^{2}$. All parameters  (9 pc resolution) are corrected by sensitivity and resolution biases by CPROPS.
\end{tablenotes}
\end{threeparttable}
\label{tab:cprops_parameters_SMC}
\end{table*}

\begin{figure*}
	\includegraphics[width=\linewidth]{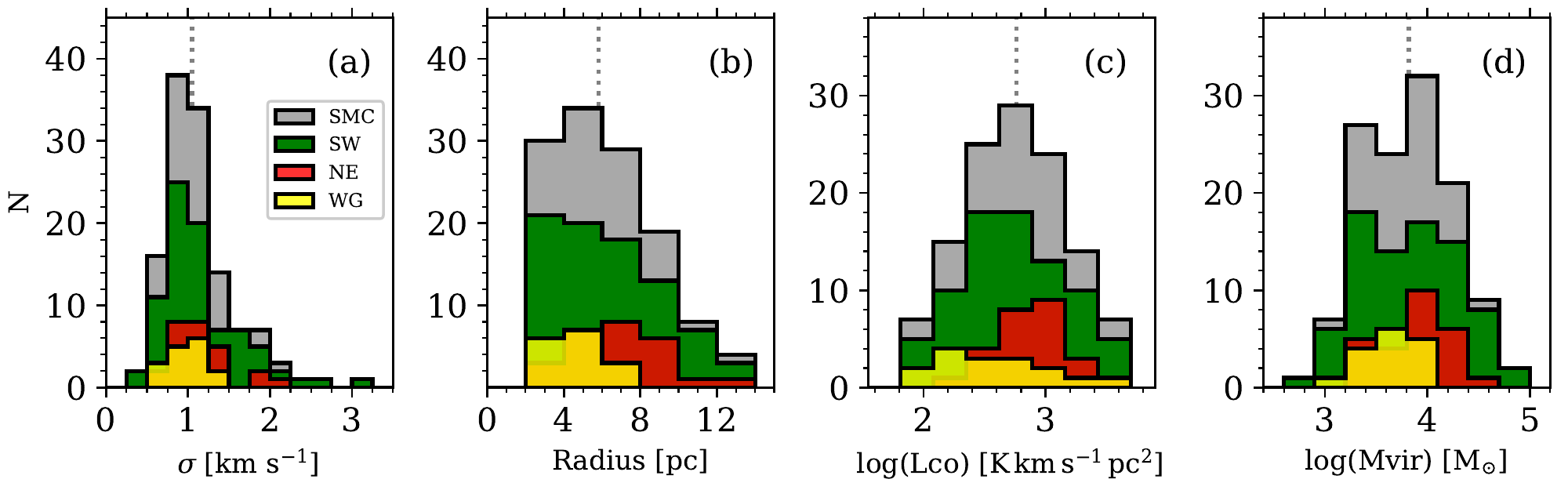}
    \caption{Histogram of the velocity dispersion $\sigma_{\upsilon}$ (a), Radius (b), CO luminosity (c), and virial mass (d) for the SMC clouds (gray histograms), and clouds from the SW (green), NE (red) and WG (yellow). These histograms belong to the well-resolved clouds with $R \gtrsim 2.2$ pc and SNRs $\gtrsim 5$. The vertical dotted line in each panel indicates the median of the parameter given in Table \ref{tab:median_parameters_smc} for the total sample. These parameters were corrected by sensitivity and resolution biases from observations at 9 pc resolution.}
    \label{fig:COparam_histo}
\end{figure*}

In Figure \ref{fig:COparam_histo}, we show the histograms of the velocity dispersion, radius, CO luminosity, and virial mass for the well-resolved, high SNR clouds (gray color). The median values of each parameter for the total sample are the following: $\sigma_{\upsilon} = 1.05\,\pm\,0.23$ \kms, $R = 5.8\,\pm\,2.0$ pc, $L_{\text{CO}} = (5.8\,\pm\,3.7)\times10^{2}$ \Lco, and $M_{\text{vir}} = (6.6\,\pm\,5.0)\times10^{3}$ \Msun\ (see Table \ref{tab:median_parameters_smc}). The distributions of the cloud parameters for the three regions, SW (green), NE (red), and WG (yellow), are also shown. Although these histograms differ in the size of the samples, the parameter distributions are centered in almost similar median values. Despite the lower number of clouds in the NE and WG regions with respect to the SW number of clouds, we quantify the comparison between the SW, NE, and WG samples by applying the Kolmogorov-Smirnov (K-S)\footnote{The K-S test is based on the "null hypothesis" which indicates that two dataset values are from the same distribution if {\it p-value} $> 10$\%.} statistical test. \textcolor{black}{\bf We find that the chance that the $R$ and $M_{\text{vir}}$ histograms of the WG clouds is the same as the ones of the NE and SW histograms 
\textcolor{black}{\bf ({\it p-value} $\sim 0.03$) } is very low.} In effect, Figure \ref{fig:COparam_histo} shows that there is a preponderance of the WG clouds to be smaller and less massive than the clouds in the other regions. Of course, a large number of clouds is required to confirm this result.

\begin{table}
\centering
\caption{Measured properties of the SMC clouds}
\begin{threeparttable}
\begin{tabular}{l c c c c c
                }
\hline
\hline
       &    \multicolumn{5}{c}{SMC}   \\
\cmidrule{2-6}
Property & Min & $25th$ & Median & $75th$ & Max \\
\hline
$\sigma_{\upsilon}$             &  0.39 &  0.85 &  1.05  &  1.31 &   3.21 \\ 
$R$                             &  2.23 &  4.06 &  5.81  &  7.98 &  13.13 \\
$\log(L_{\text{CO}})$           &  1.82 &  2.48 &  2.76  &  3.04 &   3.84  \\
$\log(M_{\text{vir}})$          &  2.86 &  3.47 &  3.82  &  4.13 &   5.04  \\
$\log(\Sigma)$                  & 0.98  &  1.63 &  1.82  &  2.02 &   2.57 \\
$\log(\sigma^{2}_{\upsilon}/R)$ & $-$1.55 & $-$0.89 & $-$0.70  & $-$0.50 & 0.05  \\
$\alpha_{\text{CO}}$            &  2.65 & 7.72 & 10.50  &  15.10 & 66.85  \\
\hline
\end{tabular}
  \begin{tablenotes}
   \item {\bf Units:} $R$ in pc; $\sigma_{\upsilon}$ in \kms;  $L_{\text{CO}}$ in \Lco; $M_{\text{vir}}$ in \Msun; $\Sigma$ in \Msun/pc$^{2}$, estimated with the $M_{\text{vir}}$; $\sigma^{2}_{\upsilon}/R$ in km$^{2}$\,s$^{-2}$\,pc$^{-1}$; $\alpha_{\text{CO}}$ in \Xco.
   \item Note: Values corresponding to well resolved clouds ($R \gtrsim 2.2$ pc and SNRs $\gtrsim 5$).  
  \end{tablenotes}
\end{threeparttable}
\label{tab:median_parameters_smc}
\end{table}

\subsection{Scaling Relations of the SMC clouds}
\label{sec:scaling_relation}

In the following, we analyze the scaling relations between the velocity dispersion, radius, luminosity, and virial mass of the SMC clouds. We determine the best fit of each relationship by using a non-parametric (bootstrapping) method. The methods use the original X and Y data of dimension $k$ and fit a power-law model of $N$ different $k$-dimensional X-Y dataset chosen randomly from the original data. The method gets $N$ slightly different models which are used to create a confidence interval for the slope and intersect. Then, from the distribution of the fitted parameters, we estimate the mean of the slope and the intersection with their respective uncertainties by means of the bootstrapping technique. These mean values will be defined as the best fit of the X-Y relation. In general, $N = 1000-2000$ is sufficient to achieve the best fit (no changes in the fitted parameter distributions) when the scatter of data is small. We apply this method to determine the correlations of the CO parameters for the well-resolved SMC clouds with SNRs $\gtrsim 5$, as well as the determination of the correlation of the parameters of the SW, NE, and WG clouds, separately. 

\subsubsection{The Size-Linewidth Relation}
\label{sec:linewidthSize_Relation}

\begin{figure*}
    \includegraphics{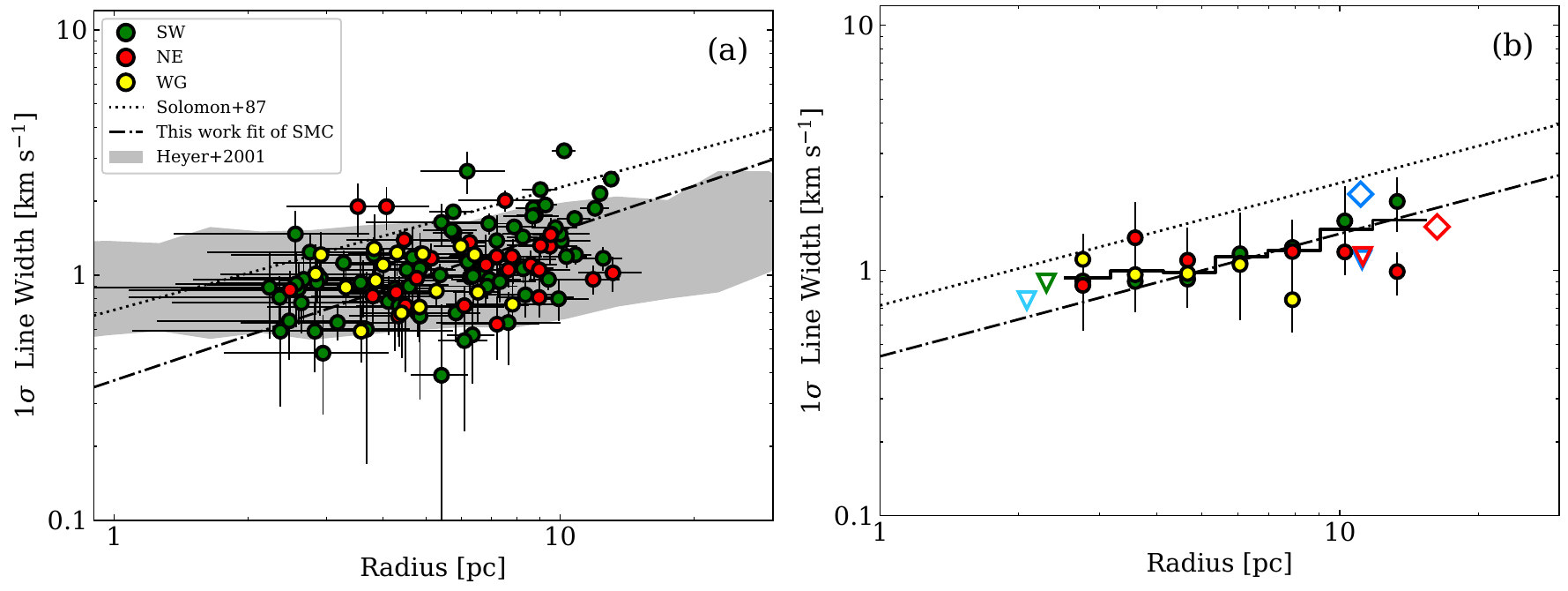}
    \caption{Size-linewidth Relation of the SMC clouds (parameters at 9 pc resolution corrected by sensitivity and resolution biases by CPROPS). The SW (SW-Bar, DarkPK) clouds are indicated in green, the NE (NE-Bar, N66) clouds in red, and the WG (N83, NGC 602, N88) clouds in yellow. In both panels, the black dotted line indicates the \cite{Solomon_1987ApJ_319_730S}'s scaling relation for Milky Way GMCs. In Panel (a), the dot-dashed line is the best fit obtained for the total sample of the SMC clouds (SNRs $>$ 5). It has a slope of $0.61\,\pm\,0.25$. The gray shadow corresponds to the location of the outer Galaxy clouds from \cite{Heyer_2001ApJ_551_852H}. Panel (b) shows the average velocity dispersion within logarithmic bins of radius (stepped line), \textcolor{black}{\bf which shows a slope (dot-dashed line) of $\sim0.5$ for $R \gtrsim 5$ pc.} The average $\sigma_{\upsilon}$ and $R$ for the NE (red triangle) and SW (red diamond) clouds by \cite{Muller_2010ApJ_712_1248M} and the average values for the SW (blue diamond) and N83 (blue triangle) clouds by \cite{Bolatto_2008ApJ_686} are shown. The green and cyan triangles correspond to median values ($\sim 1-2$ pc resolution) taken from SWBarN \citep{Saldanio_2018_BAAA_60_192S} and NGC 602  \citep{Fukui_2020arXiv200513750F}, respectively.} 
    \label{fig:sig_radius_relation}
\end{figure*}

Figure \ref{fig:sig_radius_relation}a shows the size-linewidth relation ($\sigma_{\upsilon}-R$) of the identified SMC clouds. In this figure, the SW, NE, and WG clouds are indicated with green, red, and yellow dots respectively, and the \cite{Solomon_1987ApJ_319_730S} relation for Milky Way GMCs with a dotted line. Applying the bootstrapping fit method to the $\sigma_{\upsilon}-R$ relationship (in log-log scales) to the 124 clouds with SNRs $>$ 5 in intensity, we find that the size-linewidth relation is given by: 

\begin{equation}
\label{eq:smc_linewidth_size_relation}
\begin{aligned}
 \log(\sigma_{\upsilon}) \,=\, (-0.43\,\pm\,0.18)+(0.61\,\pm\,0.25)\log(R)
\end{aligned}
\end{equation}

\noindent
with a Spearman's correlation coefficient of $\rho_{\text{S}} = 0.45$. The slope of the SMC relationship is similar to that measured by \cite{Bolatto_2008ApJ_686} for GMCs of low metallicity galaxies (including the SMC clouds observed at $6-17$ pc resolution), but our fit is slightly offset to smaller linewidths by $\sim$ 0.1 dex. The slope of the relationship increases to $\sim$ $0.8\,\pm\,0.2$,  when we include faint clouds with SNRs $>$ 3 in intensity.  This effect in the slope is identical to that found by \cite{Wong_2011ApJS_197s} for the LMC clouds showing that the decomposition of CPROPS for fainter clouds affects the size distribution resulting in steeper relations with the linewidth. This is a consequence of the considerable uncertainty introduced in the linewidth and radius by CPROPS for faint clouds, as discussed by \cite{Rosolowsky_2006PASP_118_590}. 

\textcolor{black}{\bf In panel (b) of Figure \ref{fig:sig_radius_relation}, we show the variation of the mean velocity dispersion within logarithmic bins of the radius for the total sample (stepped line), which follows the Larson law relation with a slope of $\sim 0.5$ (dot-dashed line) for $R > 5$ pc. Our average values are consistent with previous determinations at 10 pc resolution \citep[see][]{Bolatto_2008ApJ_686,Muller_2010ApJ_712_1248M}. At smaller sizes ($\lesssim 5$ pc), the SMC clouds tend to have larger average linewidths than that expected by the trend of bigger SMC clouds (dot-dashed line), consistent with findings at even smaller scales ($\sim 1-2$ pc) in the SMC \citep{Saldanio_2018_BAAA_60_192S,Muraoka_2017ApJ_844_98M,Fukui_2020arXiv200513750F}, which show that small clouds tend to have similar or larger linewidths than the inner Milky Way clouds of similar size (see Figure \ref{fig:scaling_relation_allGalaxies}a).} 
The behavior in the linewidth at different scale sizes may indicate differences in the dynamic states of small clouds and large GMCs \citep[see][]{Wong_2019ApJ_885_50W}.

\begin{figure*}
	\includegraphics{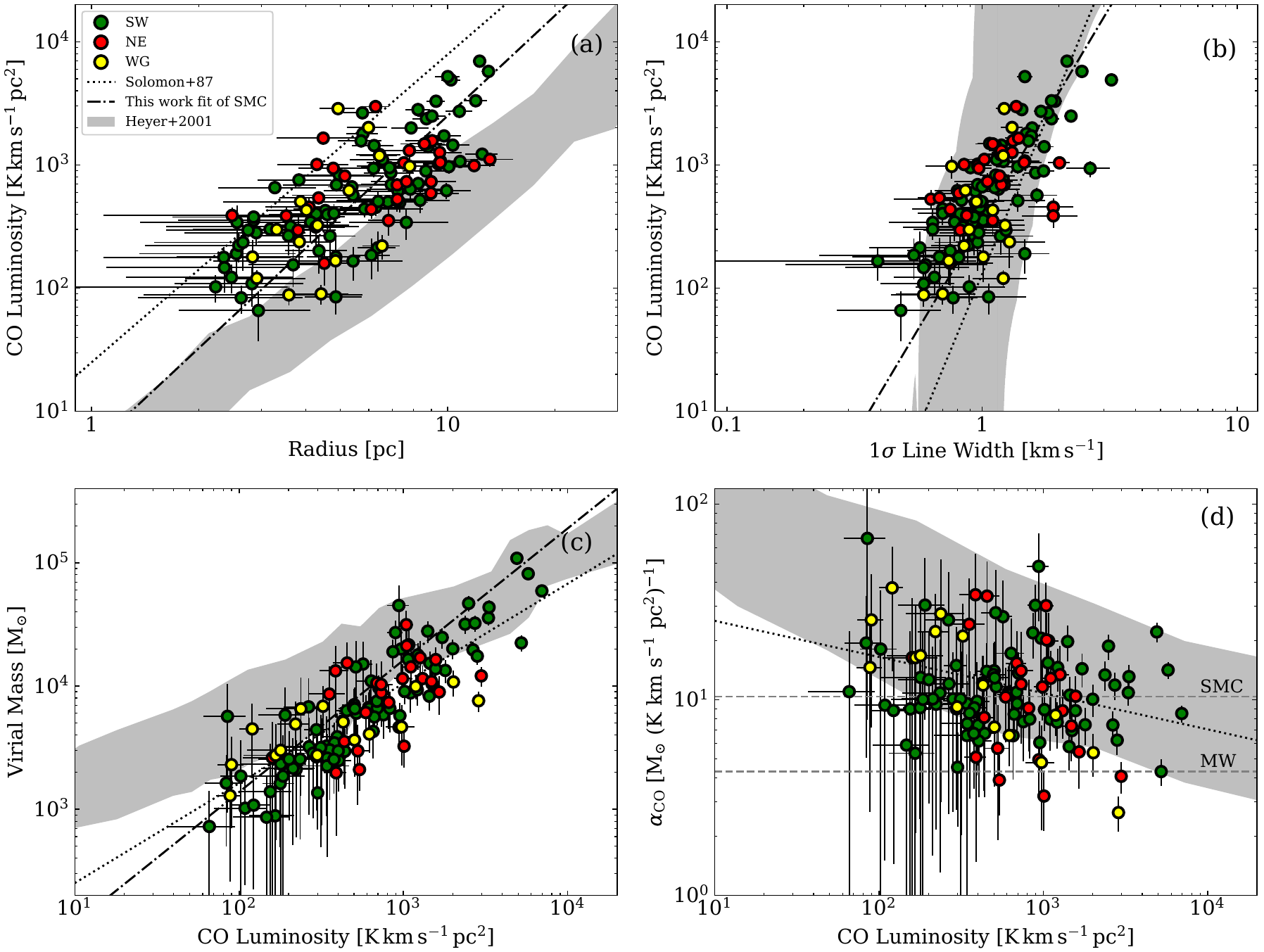}
    \caption{Scaling relation between the luminosity, radius, velocity dispersion, virial mass, and the CO-to-H$_2$ conversion factor. In all panels, the dotted lines indicate the \cite{Solomon_1987ApJ_319_730S}'s relations, while gray shadows exhibit the location of the outer Galaxy clouds \citep{Heyer_2001ApJ_551_852H}. \textcolor{black}{In panels (a), (b) and (c)} the dot-dashed lines display the best fit for the whole sample of the SMC (SNRs $> 5$ and $R > 2.2$ pc). In panel (d), the median value obtained in this work of $\alpha_{\text{CO}} = 10.5\,\pm\,5$ \Xco\ for the SMC clouds is shown with a horizontal dashed line. The canonical conversion factor ($\alpha_{\text{CO}} = 4.36$\,\Xco) for the Milky Way (MW) defined for clouds with $L_{\text{CO}} = 10^{5}$ \Lco\ is also indicated with a horizontal dashed line.}
    \label{fig:Lco_scaling_relation}
\end{figure*}

\subsubsection{The Luminosity-Scaling Relations}
\label{sec:luminosity_scaling_relation}

Panel (a) and (b) of Figure \ref{fig:Lco_scaling_relation} show the good correlation ($\rho_{\text{S}} \simeq 0.7$) of the CO luminosity with the radius and velocity dispersion for the SMC clouds. Applying the bootstrapping fitting  method to both luminosity-scaling relations, we find that the best fists are:

\begin{equation}
\label{eq:smc_Lco_size_relation}
\begin{aligned}
 \log(L_{\text{CO}}) \,=\, (0.7\,\pm\,0.2)+(2.7\,\pm\,0.3)\log(R),
\end{aligned}
\end{equation}

\noindent
and 

\begin{equation}
\label{eq:smc_Lco_sigma_relation}
\begin{aligned}
 \log(L_{\text{CO}}) \,=\, (2.67\,\pm\,0.04)+(3.6\,\pm\,0.3)\log(\sigma_{\upsilon}).
\end{aligned}
\end{equation}

\noindent
Taking into account the Milky Way luminosity-scaling relation (equations \ref{eq:lco_radius_solomon} and \ref{eq:lco-linewidth_solomon}), the SMC clouds are less luminous than the inner Milky Way clouds of similar size by a factor of $\sim$ 3. However, for similar linewidth, the luminosities of the SMC clouds tend to be higher by a factor of $\sim$ 3.5. These results are consistent with those found by \cite{Bolatto_2008ApJ_686}, who indicated that the SMC clouds tend to be underluminous for their sizes and overluminous for their velocity dispersions. Nonetheless, comparing our data with those of the outer Galaxy (gray shadow) from \cite{Heyer_2001ApJ_551_852H}, we find that the SMC clouds tend to be smaller but as turbulent as those of the outer Galaxy for a given luminosity.

Finally, we do not find considerable differences between the luminosity-scaling relations for the SW, NE, and WG clouds, which indicates that the environment in the SMC regions does not affect these scaling relations.  

\subsubsection{Virial Mass - CO Luminosity relation}
\label{sec:virial_mass_vs_Luminosity}

\textcolor{black}{ The $M_{\text{vir}}-L_{\text{CO}}$ relation of the SMC clouds is shown in Figure \ref{fig:Lco_scaling_relation}c.} The best fit of our data gives the following relation between $M_{\text{vir}}$ and $L_{\text{CO}}$:

\begin{equation}
\label{eq:smc_Mvir_Lco_relation}
\begin{aligned}
 \log(M_{\text{vir}}) \,=\, (1.0\,\pm\,0.2)+(1.0\,\pm\,0.2)\log(L_{\text{CO}}).
\end{aligned}
\end{equation}

\noindent
\textcolor{black}{ The clouds with high luminosity have similar values as those of the outer Galaxy (gray shadow) while those with lower luminosity deviate from this trend, showing lower virial masses and approaching the inner Milky Way GMCs (dotted line).} The larger size and the smaller velocity dispersion of the SMC clouds for a given luminosity, \textcolor{black}{ as can be seen in panels (a) and (b) of Figure \ref{fig:Lco_scaling_relation}, shifts in opposite directions with respect to the Milky Way clouds and approximately cancel out when the virial mass is calculated, giving a strong correlation between $M_{\text{vir}}$ and $L_{\text{CO}}$ ($\rho_{\text{S}} \simeq 0.85$) and a good agreement among the SMC and inner Milky Way clouds between $10^{3}$ to $10^{4}$ \Msun. The deviation seen between} the SMC and outer Galaxy clouds for lower luminosities may be explained by a selection effect in the \cite{Heyer_2001ApJ_551_852H}'s analysis, who excluded the narrow-line regions from the cloud catalog. 

The $M_{\text{vir}}-L_{\text{CO}}$ correlation found for the SMC clouds (equation \ref{eq:smc_Mvir_Lco_relation}) with a slope of $\sim 1.0\,\pm\,0.2$ implies a roughly constant virial-based CO-to-H$_2$ conversion factor ($\alpha_{\text{CO}} = M_{\text{vir}}/L_{\text{CO}}$) for a luminosity range between $60$ to $5\times10^{3}$ \Lco. In panel (d) of Figure \ref{fig:Lco_scaling_relation}, we show that the  $\alpha_{\text{CO}}-L_{\text{CO}}$ relationship of the SMC clouds fall within the trend found for the outer Galaxy clouds (gray shadow) which follow $\alpha_{\text{CO}} \propto L_{\text{CO}}^{-0.5}$ \citep{Heyer_2001ApJ_551_852H}. The Milky Way clouds, following  $\alpha_{\text{CO}} \propto L_{\text{CO}}^{-0.2}$, is shown with dotted line in the figure. For the SMC, we find a slope of $-0.20\,\pm\,0.17$ with a Spearman's coefficient of $\rho_{\text{S}} = -0.2$ implying a very weak anti-correlation trend.

\textcolor{black}{ We determine the virial-based CO-to-H$_2$ conversion factor for the SMC of $\alpha_{\text{CO(1-0)}} = 10.5$ \Msun(\Lco)$^{-1}$, with a standard deviation of $\sim 5$ \Msun(\Lco)$^{-1}$. This $\alpha_{\rm CO}$ is the median value of the conversion factors of the 124 CO clouds using their virial masses and CO luminosities of Table \ref{tab:cprops_parameters_SMC}. } This value is $\sim 2.5$ times larger than the canonical Galactic value, \textcolor{black}{which is obtained for clouds with typical $L_{\text{CO}} = 10^{5}$ \Lco.} Our derived SMC value is consistent with the expectation of a higher $\alpha_{\text{CO}}$ for Galactic virialized clouds with $L_{\text{CO}} \sim 2\times10^{3}$ \Lco\ \citep{Solomon_1987ApJ_319_730S,Bolatto_2013ARA&A_51}. We do not find differences between the median conversion factor estimated for the SW, NE, and WG clouds to that determined for the total sample.

\subsection{Association of CO clouds with YSOs and HII Regions}
\label{sec:CO_star-forming}

We inspect the association of the CO clouds with YSOs and HII regions to assess the CO cloud parameters in star-forming regions. We plot the YSO and HII region positions superimposed on the CO distribution for all studied regions. Details of the comparison of CO emission with star-formation tracers are given in Appendix \ref{app:CO_yso_HII}.

For the YSO-CO cloud association, we use the positions of more than 1000 YSOs identified in the SMC \citep{Simon_2007ApJ_669_327S,Carlson_2011ApJ_730_78C,Sewilo_2013ApJ_778_15S}. Here, we require that the position differences between YSOs and CO clouds must be less than the deconvolved radius of the clouds\footnote{For unresolved clouds without a measured size, a radius of 2.2 pc (the limit of resolved clouds) are assigned.}. With this criteria, we find that {\bf $\sim 43$\% ($76/177$)} of the CO clouds may be associated with YSOs. We also associate the CO clouds with the HII regions using the 215 HII regions in the main body of the SMC identified by \cite{Pellegrini_2012ApJ_755_40P}. In this case, we compare the HII position within twice the CO radius as the HII region may have expanding shells that could affect the nearby environment of the molecular cloud. We find that $\sim$24\% (42/177) of the SMC clouds are spatially associated with HII regions in our observed areas.
\textcolor{black}{\bf Our analysis also shows that after accounting for the fact that there is some overlap (16\% of the clouds are associated with both YSOs and HII regions), about half of the clouds are associated with ongoing star formation (either YSOs or HII regions), but also close to half the clouds are not. This is consistent with the findings in other nearby galaxies. \citep[e.g.,][]{Corbelli_2017_AA_601A}.}

\begin{figure*}
    \includegraphics[width=\linewidth]{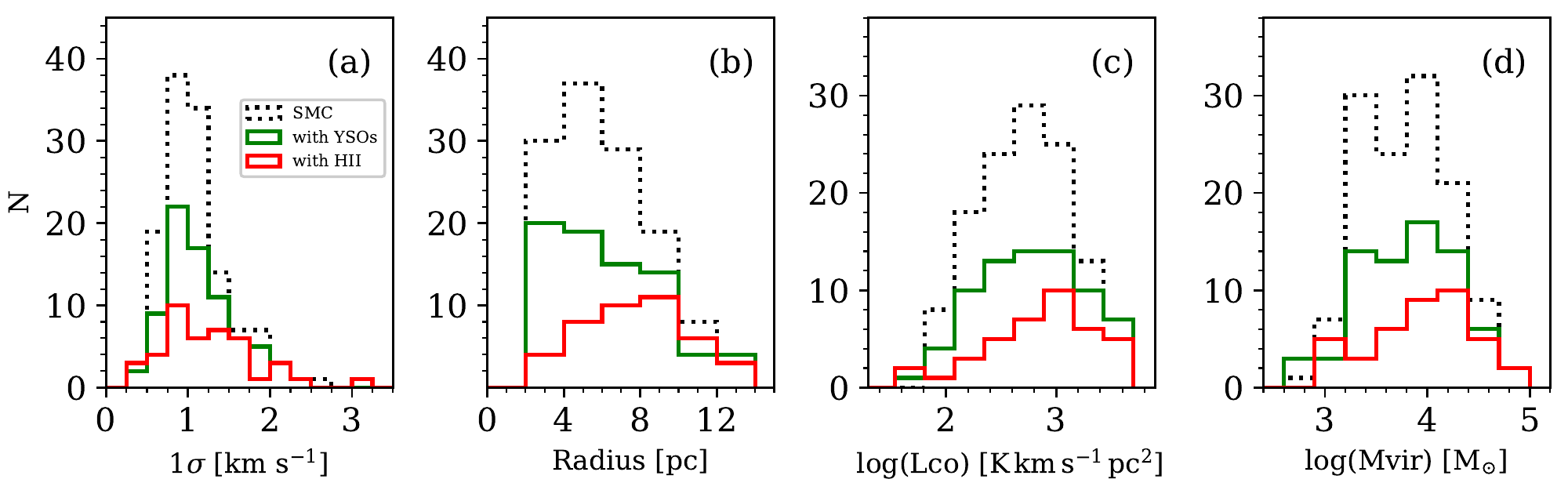}
    \caption{Histogram of the velocity dispersion (a), Radius (b), CO luminosity (c), and virial mass (d) of those CO clouds of the SMC spatially associated with YSOs (green) and HII regions (red). The histograms in dotted lines correspond to the full sample of the SMC (gray histograms in Figure \ref{fig:COparam_histo}).}    \label{fig:CO-YSO-HII_histogram}
\end{figure*}

With the resulting associations, we constructed histograms of the molecular parameters of CO clouds likely associated with YSOs (green color) and HII regions (red color) shown in Figure \ref{fig:CO-YSO-HII_histogram}. We plot the histogram of the SMC total sample (dotted line) for comparison. To determine quantitatively an association, we performed a K-S test on the SMC total sample and those clouds associated with YSOs and HII regions. \textcolor{black}{The K-S result shows that the SMC clouds associated with YSOs have similar velocity dispersion, radius, CO luminosity, and virial mass as to the total CO sample of the SMC ({\it p-values} $> 30$\%), therefore they are consistent with belonging to the same parent distribution. 
However, the histograms of the CO clouds associated with HII regions and the total SMC sample yield {\it p-values} $\lesssim 5$\%, suggesting that HII regions are preferentially associated with larger, more luminous, and more massive clouds (Figure \ref{fig:CO-YSO-HII_histogram}b, c,d).} 

\subsection{The Mass Distribution of the SMC clouds}
\label{sec:mass_spectra}

For the first time, we have a large number (over 100)  of resolved clouds to determine the mass spectrum of the SMC molecular clouds. The mass distribution is used to quantify the characteristics of the  populations between systems and gives important parameters for theories and models of the formation of star-forming clouds. Instead of using the power-law mass spectrum ($dN/dM \propto M^{\beta}$), we  use the integration of this mass spectrum known as the cumulative mass distribution (CMD), given by a power-law distribution as:

\begin{equation}
\label{eq:CMD}
\begin{aligned}
  N(M'\,>\,M) = \left(\,\frac{M}{M_{0}}\,\right)^{\,\beta +1}.
\end{aligned}
\end{equation}

\noindent
This distribution gives the number of clouds ($N$) with masses ($M'$) greater than a reference mass ($M$) as a function of that reference mass. In all known cases, the power-law index $\beta$ takes values lower than -1. In general, in cases where $\beta\,>\,-2$, the total galaxy's molecular gas is dominated by massive GMCs, while for steeper distributions with $\beta\,<\,-2$, most of the total molecular mass of a galaxy is contained in smaller clouds. $M_0$ is the maximum mass in the distribution and generally scales with the total mass of the sample.

The power-law distribution usually tends to over-predict the number of clouds at very high masses where the mass spectrum can undergo a sharp roll-off with a truncated end at $M_0$. It has been found that this feature is very pronounced in the mass spectrum of the inner disk Milky Way \citep{Rosolowsky_2005PASP_117_1403R}, in the galactic center of M51 \citep{Colombo_2014ApJ_784_3C}, and M33 \citep{Gratier_2012_AA_542A_108G}. To take this feature into account we consider the truncated CMD, given by:

\begin{equation}
\label{eq:truncated_CMD}
\begin{aligned}
  N(M'\,>\,M) = N_{0} \left[\left(\,\frac{M}{M_{0}}\,\right)^{\,\beta +1}\, - \,1\right],
\end{aligned}
\end{equation}

\noindent
In this distribution, $N_0$ is the number of clouds with masses $>$\,$2^{1/(\beta\,+\,1)}M_{0}$ where the CMD begins to deviate significantly from the power law. For $N_0\,\sim\,1$, the deviation is negligible. 

\begin{figure*}
    \centering
    \includegraphics{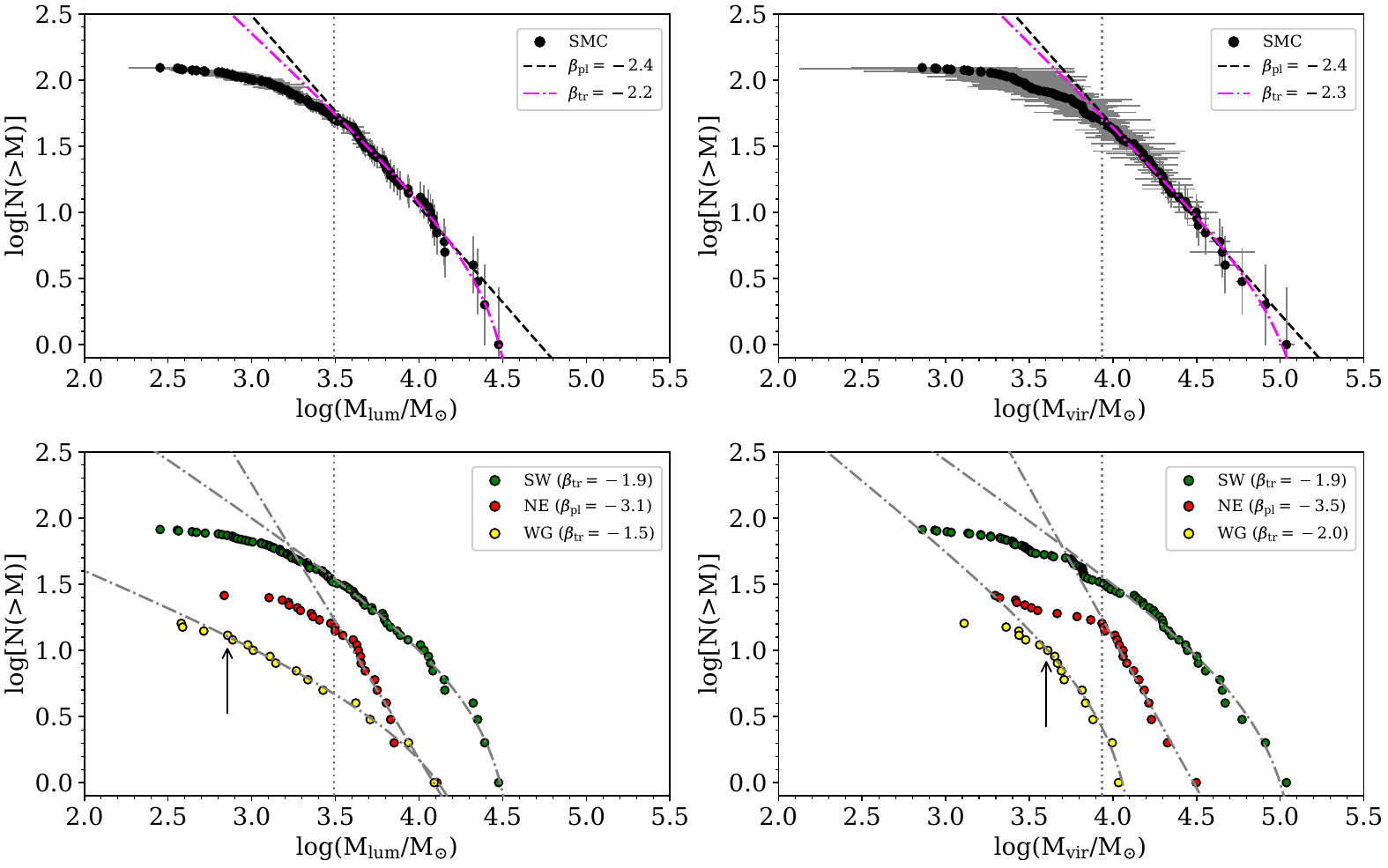}
    \caption{Top Panels: Cumulative luminous mass distribution (top left) and cumulative virial mass distribution (top right) of the SMC clouds shown in black dots (at 9 pc resolution). In gray bars, the error in both axes is indicated. The vertical dotted line corresponds to the completeness limits ($M_{\text{lum}}\,=\,3.1\,\times\,10^{3}$\,\,\Msun, and $M_{\text{vir}}\,=\,8.6\,\times\,10^{3}$\,\,\Msun) from which the best fit is applied to the SMC spectra. The power law model best fit ($\beta_{\rm pl} = -2.4$) is shown in black dashed lines, while the truncated model ($\beta_{\rm tr} \simeq -2.2$) best fit  is shown with magenta dot-dashed lines. Bottom Panels: The CMD of the SW, NE, and WG clouds are indicated in green, red, and yellow dots, respectively. In gray dot-dashed lines, the best fit of these CMDs is indicated. In the case of the WG clouds, the black arrow shows that includes more clouds in both mass distributions for a completeness limit of $M_{\text{lum}}\,=\,7.15\,\times\,10^{2}$\,\,\Msun\, and $M_{\text{vir}}\,=\,4.0\,\times\,10^{3}$\,\,\Msun.
    } 
    \label{fig:mass_spectrum}
\end{figure*}

The advantage of using the CMD as a mass spectrum is that we can generate the mass distribution without choosing a bin size, which introduces some bias in the estimation of the main parameters of equations \ref{eq:CMD} and \ref{eq:truncated_CMD} \citep{Rosolowsky_2005PASP_117_1403R}. We use the methodology described in \cite{Rosolowsky_2005PASP_117_1403R} by using the IDL program provided by this author (private communication). The algorithm uses the ``error-in-variables" method, which maximizes the likelihood that a set of data $\{M$, $N\}$, with associated uncertainties, can be modeled with parameters $\{N_0$, $M_0$, $\beta\}$. We determine these parameters for the luminous mass distribution ($M_{\text{lum}}$ from equation \ref{eq:luminous_mass}), and for the virial mass distribution ($M_{\text{vir}}$ from equation \ref{eq:virial_mass}). In Figure \ref{fig:mass_spectrum}, the CMD for both luminous and virial mass of the 124 SMC clouds (SNR $> 5$ and $R > 2.2$ pc) are shown in black dots. The fit of the cumulative luminous mass distribution is performed above the completeness limit of $M_{\text{lum}}\,\simeq\,3.1\,\times\,10^{3}$\,\,\Msun, while for the cumulative virial mass distribution the completeness limit is $M_{\text{vir}}\,\simeq\,9.0\,\times\,10^{3}$\,\,\Msun\ (gray vertical dashed line in both panels). Both luminous and virial mass distribution includes 50 clouds above the completeness limits. In Table \ref{tab:CMD_bestFits2}, the parameters of the best fit of the power-law and the truncated models, \textcolor{black}{and the number of clouds used for the fit are indicated.}
\textcolor{black}{We also included in Table 4 the best fits for the three regions (SW, NE, and WG) which are included in Figure \ref{fig:mass_spectrum}. As they have very low numbers of clouds and thus suffer from uncertainties in the CMDs, we have not included these fits  in the following analysis.}

\begin{table}
\centering
\caption{Best fit of the CMDs for the SMC clouds.}
\begin{threeparttable}
\begin{tabular}{llcccc}
\hline
\hline
\multicolumn{6}{c}{Luminous Mass Distribution} \\
\hline
Region  & Model &  $\beta$ &      $M_0$        & $N_0$  & N\tnote{a} \\
        &       &          & ($10^{4}$\,\Msun) &        &   \\
\hline
SMC & PLaw   & $-2.4\pm0.2$ & \,\,\,$5.3\pm1.5$ &     $-$     &  50 \\
SMC & Trunc. & $-2.2\pm0.3$ & \,\,\,$3.8\pm0.9$ & $3.0\pm3.0$ & 50  \\
SW  & Trunc. & $-1.9\pm0.3$ & \,\,\,$3.8\pm0.8$ & $4.5\pm4.0$ & 32  \\
NE  & PLaw   & $-3.1\pm0.6$ & \,\,\,$1.2\pm0.3$ &     $-$     & 15 \\
WG\tnote{b}  & Trunc. & $-1.5\pm0.3$ & \,\,\,$2.6\pm1.3$ & $2.3\pm2.0$ & 13  \\
\hline
\hline
\multicolumn{5}{c}{Virial Mass Distribution} \\
\hline
Region  & Model &  $\beta$ &      $M_0$        & $N_0$  & N\tnote{a} \\
        &       &          & ($10^{4}$\,\Msun) &        &   \\
\hline
SMC & PLaw   & $-2.4\pm0.2$ &      $14.5\pm4.2$ &     $-$     & 50 \\
SMC & Trunc. & $-2.3\pm0.2$ &      $15.5\pm5.8$ & $1.4\pm1.0$ & 50 \\
SW  & Trunc. & $-1.9\pm0.3$ &      $13.3\pm4.0$ & $3.4\pm3.0$ & 32 \\
NE  & PLaw   & $-3.3\pm0.5$ & \,\,\,$3.0\pm2.1$ &     $-$     & 15 \\
WG\tnote{b}  & Trunc. & $-2.0\pm0.2$ & \,\,\,$1.4\pm1.0$ & $3.7\pm2.0$ & 10 \\
\hline
\end{tabular}
  \begin{tablenotes}
   \item {\bf Note:} The best fit of the Power Law (PLaw) and Truncated (Trunc.) models for the SMC clouds are given. In the case of the SW, NE, and WG regions only the models which best fit the CMDs are indicated. The fitting is performed above the completeness limit of $M_{\text{lum}} = 3.1\times10^{3}$ \Msun\ and $ M_{\text{vir}} = 8.6\times10^{3}$ \Msun\ for all regions. 
   \item[a] The last column indicates the number of well-resolved clouds with SNR\,\,$> 5$ used for the fit.
   \item[b] For the WG region, the completeness limits are  $M_{\text{lum}} = 7.15\times10^{2}$ \Msun\ and $ M_{\text{vir}} = 4\times10^{3}$ \Msun.\\ 
  \end{tablenotes}
\end{threeparttable}
\label{tab:CMD_bestFits2}
\end{table}

The best fit for the SMC mass spectra gives power indexes $\beta\,<\,-2$. This indicates that the molecular gas in the SMC is preferentially distributed in low-mass clouds. The cumulative luminous mass distribution of the SMC shows a moderated truncation in the high-mass end with $N_0\,=\,3.0\,\pm\,3.0$ and a maximum mass of $M_0\,=\,(3.8\,\pm\,0.9)\,\times\,10^{4}$ \Msun, while for the cumulative virial mass distribution the truncation is negligible with $N_0\,=\,1.3\,\pm\,1.0$, and $M_0\,=\,(15.8\,\pm\,5.0)\,\times\,10^{4}$ \Msun. Therefore, the cumulative virial mass distribution follows a power-law distribution (within the errors) beyond the completeness limit rather than a truncated model. We find that the CMD for the SMC is similar to those obtained in the nearest irregular galaxy LMC by \cite{Wong_2011ApJS_197s}, except that the maximum mass $M_0$ in both luminous and virial mass distributions for the SMC are about one order of magnitude lower than those of the LMC.

The power indexes of the CMDs found in the SMC are also generally consistent with those estimated for galaxies with non-spiral morphology like the irregular LMC galaxy \citep[$\beta \simeq -2.3$,][]{Wong_2011ApJS_197s}, the lenticular galaxy NGC 4526 \citep[$\beta \simeq -2.4$,][]{Utomo_2015ApJ_803}, or in more quiescent environments beyond the galactic center of spiral galaxies like M33 \citep[$\beta =  -2.3$,][]{Gratier_2012_AA_542A_108G}, M51 \citep[$\beta \simeq -2.5$,][]{Colombo_2014ApJ_784_3C} and the outer Milky Way \citep[$\beta = -2.5$,][]{Rosolowsky_2005PASP_117_1403R}, whereas shallower CMDs ($\beta > -2.0$) are found towards the center of these galaxies \citep{Rosolowsky_2005PASP_117_1403R,Braine_2018_AA_612A_51B}.

As we will see in Section \ref{sec:discussion}, the SMC clouds tend to be in virial equilibrium so that the virial mass values represent the total gas mass of the clouds, suggesting that the virial mass distribution would be a good representation of the real CMD of the SMC. In this scenario, the SMC would be dominated by low-mass clouds across the galaxy (within uncertainties) unlike spiral galaxies.

\section{Discussion}
\label{sec:discussion}

\subsection{The stability of the SMC clouds}
\label{sec:stability_SMC}

\begin{figure*}
    \centering
	\includegraphics{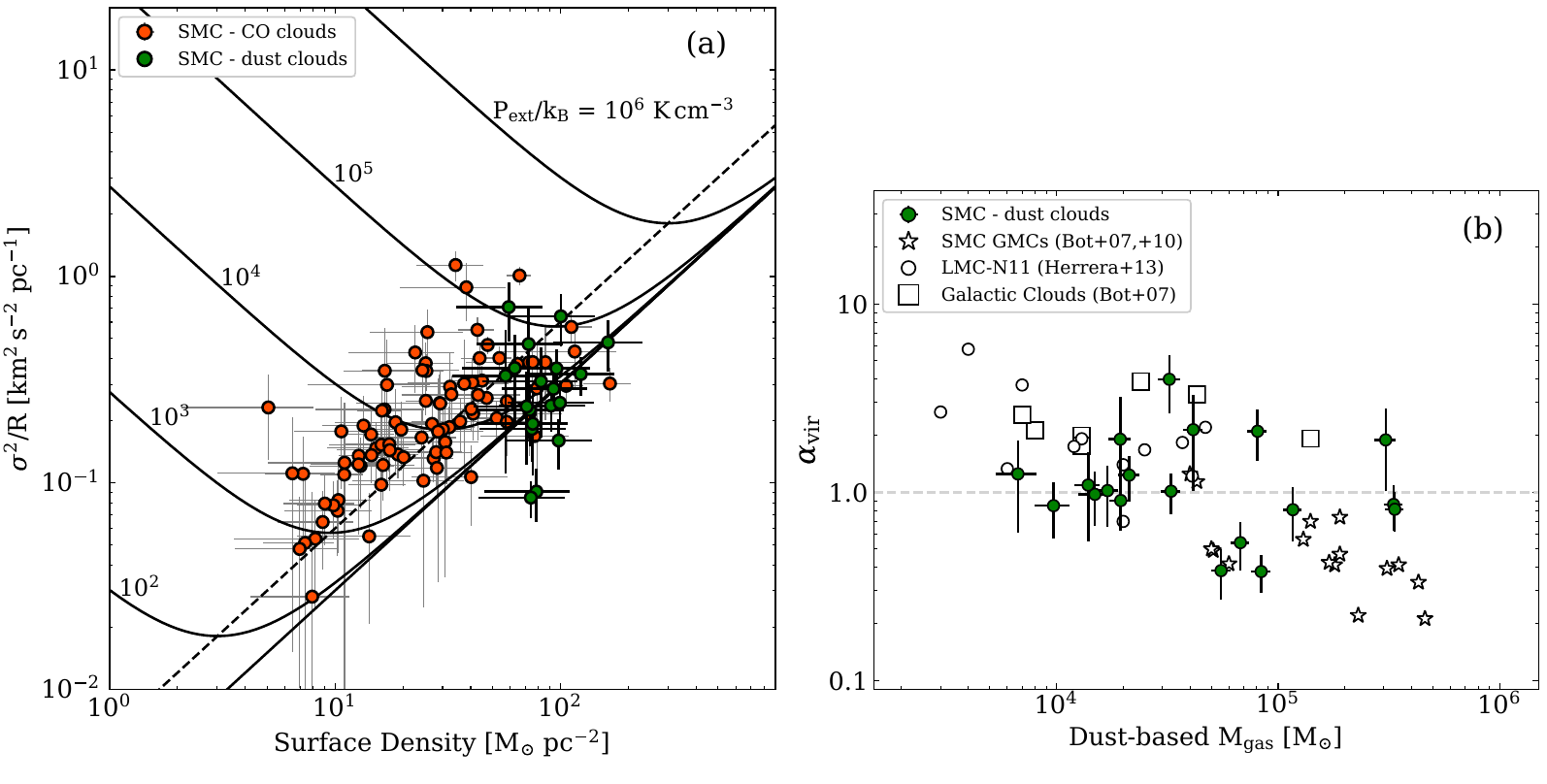}
    \caption{\bf Panel (a): $\Sigma - \sigma_{\upsilon}^{2}/R$ relationship for the SMC clouds. The red dots show the CO clouds of this work, whose $\Sigma$ is calculated from their luminous mass assuming the Galactic conversion factor. The green dots correspond to the dust clouds identified by \cite{Takekoshi_2017ApJ_835_55T} at 12 pc resolution, in which their $\Sigma$ are calculated from the dust-based total gas mass estimated by these authors and the velocity dispersion by mean equation \ref{eq:equivalent_linewidth}. The solid black lines indicate gravitational virial conditions for clouds confined by external pressures from $P_{\text{ext}}/k_{\text{B}} = 0$ to $10^{6}$ K\,\cmc. The dashed line corresponds to the marginal gravitationally bound condition. Panel (a): Virial parameter ($\alpha_{\text{vir}}$) vs. dust-based total gas mass for the same SMC dust clouds shown in panel (a). We have added for comparison the virial parameters of the Milky Way (squares) and SMC GMCs (star symbols) \citep{Bot_2007AA_471_103B,Bot_2010AA_524A52B}, and the LMC GMCs (empty circles) from \cite{Herrera_2013AA_554A_91H}. The average virial parameter for the SMC dust clouds is $\alpha_{\text{vir}} = 1.1^{+1.2}_{-0.3}$.} 
    \label{fig:Mvir-Mtot_relation}
\end{figure*}

\textcolor{black}{\bf We analyze the dynamical state of the SMC clouds plotting the $\sigma_{\upsilon}^{2}/R$ ratio as function of the surface density $\Sigma = M_{\text{gas}}/\pi\,R^{2}$, where $\sigma_{\upsilon}$ is the velocity dispersion, $R$ the (deconvolved) radius and $M_{\rm gas}$ the total gas mass of the cloud. This relationship is shown in panel (a) of figure \ref{fig:Mvir-Mtot_relation}, where the black curves indicate clouds in virial equilibrium with an external pressure \citep{Field2011}. 
The straight solid line corresponds to virialized clouds without external pressure, and the straight dashed line indicates marginal gravitationally bound conditions (identical kinetic and gravitational energy) with $P_{\rm ext} = 0$.}

\textcolor{black}{\bf We show in Figure \ref{fig:Mvir-Mtot_relation}a our results for the 124 well-resolved and high SNR CO clouds (identified in this work) in red dots. We place these clouds in the diagram using the Galactic conversion factor to estimate their surface densities.
Their $\Sigma$ range from $10$ to $100$ \Msun\,pc$^{-2}$. For the assumed conversion factor,  most of the CO clouds are close to the marginally gravitationally bound line. If we use a larger conversion factor, as is expected in low metallicity galaxies \citep{Leroy_2011ApJ_737_12L,Jameson_2018_ApJ_853_111J}, the luminous CO mass would increase and the CO clouds would move to the right in the plot, becoming more strongly bound.}
 
\textcolor{black}{\bf For an alternative result independent of conversion factor assumptions, we also examine the stability of the molecular clouds identified by \cite{Takekoshi_2017ApJ_835_55T} in the 1.1 mm survey of the SMC at 12 pc resolution  using their estimated dust-based mass. \citeauthor{Takekoshi_2017ApJ_835_55T} estimates cloud gas masses by fitting the far-IR spectral energy distribution (SED) with a modified black body model using a dust emissivity index $\beta=1.2$ and GDR$=1000$. Within our observed area there are 19 dust clouds with reliable deconvolved radii in the range $\sim 5-40$ pc \cite[see Table 2 of][]{Takekoshi_2017ApJ_835_55T}.} 

\textcolor{black}{\bf To estimate the velocity dispersion of these 19 clouds, we use an "equivalent linewidth" ($\sigma_{\upsilon,\text{eq}}$) following the method by \cite{Heyer_2001ApJ_551_852H}, }

\begin{equation}
  \sigma_{\upsilon,\text{eq}} = \frac{I_{\text{CO}}}{\sqrt{2\pi}\,T_{\text{peak}}},
  \label{eq:equivalent_linewidth}
\end{equation}

\noindent
\textcolor{black}{\bf where $I_{\text{CO}}$ is the CO intensity and  $T_{\text{peak}}$ the peak temperature measure within the dust clouds. We determine both $I_{\text{CO}}$ and $T_{\text{peak}}$ by convolving the CO APEX of 30" (9 pc) resolution to the 1.1 mm data resolution of 40" (12 pc). Then we obtain the integrated CO spectrum over the area of the dust cloud as defined by \cite{Takekoshi_2017ApJ_835_55T}. We use this method because it is less sensitive to broad line wings or the presence of multiple CO cloud components along the line of sight than a moment-based measurement, and it requires no assumption about the line shape \citep[unlike fitting the integrated CO spectrum,][]{Leroy_2016_ApJ_831_16L}. We find $\sigma_{\upsilon,\text{eq}}$ is roughly similar to the line width of the brightest CO component in the line of sight.}

\textcolor{black}{\bf As Figure \ref{fig:Mvir-Mtot_relation}a shows, most of the dust-identified clouds (green dots) are close to the gravitational virial equilibrium (straight solid line). The dynamic range for $\Sigma$ is small, from $\sim 50$ to $100$ \Msun\,pc$^{-2}$, but this is likely a limitation of the dust observations. We list the parameters of the dust clouds in Table \ref{tab:SMC_dust_paramters}.}

\textcolor{black}{\bf In Figure \ref{fig:Mvir-Mtot_relation}b we show the virial parameter ($\alpha_{\text{vir}}$) for the 19 dust-identified SMC clouds as function of their gas mass ($M_{\text{gas}}$). We compute the virial parameter as $\alpha_{\text{vir}}=M_{\rm vir}/M_{\rm gas}$, where $M_{\rm vir}$ is from equation \ref{eq:virial_mass} using the "equivalent linewidth" and the deconvolved radius, while $M_{\rm gas}$ is from \cite{Takekoshi_2017ApJ_835_55T}. We measure an  $\alpha_{\text{vir}}$ within $\pm0.3$~dex of unity for most clouds, although the full range is $\alpha_{\text{vir}}\sim 0.4-4$. The average virial parameter for these SMC clouds is $\alpha_{\rm vir} = 1.1_{-0.3}^{+1.2}$, in very good agreement with virial equilibrium within the errors. This suggests that any support by a magnetic field plays a secondary role in the cloud equilibrium, in good agreement with recent measurements of SMC magnetic fields \citep{Lobo_Gomes_2015ApJ_806_94L,Kaczmarek_2017_MNRAS_467_1776K} which are too low to provide significant support  \citep{Bot_2007AA_471_103B}.}

\begin{table*}
\centering
\caption{Physical parameters for SMC dust clouds from \cite{Takekoshi_2017ApJ_835_55T} within our observed regions.}
\begin{threeparttable}
\begin{tabular}{l
                S[table-format=1.1]@{\,\( \pm \)\,}
                S[table-format=1.1]
                S[table-format=2.1]@{\,\( \pm \)\,}
                S[table-format=1.1]
                S[table-format=2.1]@{\,\( \pm \)\,}
                S[table-format=2.1]                
                S[table-format=3.1]@{\,\( \pm \)\,}
                S[table-format=3.1]
                S[table-format=3.1]@{\,\( \pm \)\,}
                S[table-format=2.1]
                S[table-format=2.1]@{\,\(\pm\)\,}
                S[table-format=2.1]
                S[table-format=1.1]@{\,\(\pm\)\,}
                S[table-format=1.1]
                S[table-format=3.1]@{\,\( \pm \)\,}
                S[table-format=2.1]}
\hline
\hline
  ID\tnote{a}  &    
  \multicolumn{2}{c}{$\sigma_{\upsilon}$\tnote{b}}   &
  \multicolumn{2}{c}{$R_{\text{deconv}}$\tnote{c}}                   &
  \multicolumn{2}{c}{$L_{\text{CO}}$\tnote{d}}       &
  \multicolumn{2}{c}{$M_{\text{vir}}$}      &
  \multicolumn{2}{c}{$M_{\text{gas}}$\tnote{c}}      &
  \multicolumn{2}{c}{$\Sigma_{\text{gas}}$} &
  \multicolumn{2}{c}{$\alpha_{\text{vir}}$\tnote{e}} &
  \multicolumn{2}{c}{$\alpha_{\text{CO}}$\tnote{f}} \\
      &  
      \multicolumn{2}{c}{(\kms)}             &
      \multicolumn{2}{c}{(pc)}               &
      \multicolumn{2}{c}{(caption)}          & 
      \multicolumn{2}{c}{(10$^{3}$\Msun)}    &
      \multicolumn{2}{c}{(10$^{3}$\Msun)}    &
      \multicolumn{2}{c}{(\Msun\,pc$^{-2}$)} &
      \multicolumn{2}{c}{$-$}                  &
      \multicolumn{2}{c}{(caption)}    \\
\hline
SW-1  & 2.8 & 0.2 & 34.1 & 6.8 & 111.0 & 45.1 & 278.0 &  68.2 & 332.1 & 32.7 &   90.9 &   37.3 & 0.8 & 0.2 &   29.9 &   12.5 \\
SW-2  & 2.8 & 0.1 & 32.8 & 6.6 &  89.4 & 35.8 & 267.4 &  57.1 & 336.1 & 33.2 &   99.4 &   41.2 & 0.8 & 0.2 &   37.6 &   15.5 \\
SW-3  & 3.2 & 0.3 & 16.0 & 3.2 &  56.7 & 23.3 & 170.4 &  46.7 &  80.8 &  8.0 &  100.5 &   41.4 & 2.1 & 0.6 &   14.3 &    6.0 \\
SW-4  & 1.5 & 0.1 & 14.8 & 3.0 &  47.1 & 19.3 &  34.6 &   8.4 &  67.6 &  6.7 &   98.2 &   41.0 & 0.5 & 0.1 &   14.4 &    6.1 \\
SW-5  & 1.2 & 0.1 & 15.0 & 3.0 &   8.3 &  3.3 &  22.5 &   5.9 &  55.3 &  5.5 &   78.2 &   32.2 & 0.4 & 0.1 &   66.6 &   27.3 \\
SW-6  & 1.4 & 0.2 &  8.6 & 1.7 &   5.7 &  2.3 &  17.5 &   6.1 &  17.0 &  2.0 &   73.2 &   30.2 & 1.0 & 0.4 &   29.8 &   12.5 \\
SW-7  & 1.7 & 0.1 & 10.6 & 2.1 &  14.8 &  5.9 &  31.9 &   7.4 &  32.9 &  3.4 &   93.2 &   38.2 & 1.0 & 0.2 &   22.2 &    9.2 \\
SW-9  & 2.5 & 0.6 & 13.5 & 2.7 &   9.7 &  4.1 &  87.8 &  45.7 &  41.4 &  4.1 &   72.3 &   29.8 & 2.1 & 1.1 &   42.7 &   18.5 \\
SW-20 & 1.6 & 0.2 &  5.4 & 1.1 &   5.1 &  2.0 &  14.4 &   4.6 &  14.9 &  2.3 &  162.6 &   70.9 & 1.0 & 0.3 &   29.2 &   12.3 \\
SW-22 & 1.4 & 0.3 &  7.9 & 1.6 &   4.5 &  2.0 &  16.1 &   7.6 &  13.9 &  1.8 &   70.9 &   30.1 & 1.2 & 0.6 &   30.9 &   14.3 \\
NE-1  & 3.8 & 0.8 & 39.4 & 7.9 &  80.1 & 33.9 & 591.7 & 275.9 & 305.9 & 33.7 &   62.7 &   26.1 & 1.9 & 0.9 &   38.2 &   16.7 \\
NE-2  & 3.1 & 0.4 & 13.2 & 2.6 &  15.0 &  6.1 & 131.9 &  42.8 &  32.3 &  3.7 &   59.0 &   24.2 & 4.1 & 1.4 &   21.5 &    9.1 \\
NE-3  & 1.7 & 0.1 &  8.4 & 1.7 &  10.6 &  4.2 &  25.2 &   5.9 &  21.3 &  2.4 &   96.1 &   40.4 & 1.2 & 0.3 &   20.1 &    8.3 \\
NE-4  & 1.9 & 0.6 & 10.4 & 2.1 &   3.2 &  1.4 &  39.0 &  25.9 &  19.4 &  2.3 &   57.1 &   24.0 & 2.0 & 1.4 &   60.6 &   27.5 \\
NE-6  & 2.0 & 0.2 & 22.3 & 4.5 &  32.9 & 13.3 &  92.8 &  26.4 & 116.6 & 11.2 &   74.6 &   31.0 & 0.8 & 0.2 &   35.4 &   14.7 \\
NE-12 & 1.3 & 0.3 &  5.1 & 1.0 &   3.0 &  1.3 &   9.0 &   4.5 &   6.7 &  1.4 &   82.0 &   36.4 & 1.3 & 0.7 &   22.3 &   10.7 \\
WG-2  & 1.3 & 0.0 & 19.0 & 3.8 &  38.8 & 15.5 &  33.4 &   6.8 &  83.9 &  8.7 &   74.0 &   30.6 & 0.4 & 0.1 &   21.6 &    8.9 \\
WG-4  & 1.5 & 0.1 &  7.1 & 1.4 &   8.7 &  3.5 &  16.6 &   3.5 &  19.5 &  2.2 &  123.1 &   50.5 & 0.9 & 0.2 &   22.4 &    9.4 \\
N88   & 1.1 & 0.2 &  6.4 & 1.3 &   6.4 &  2.7 &   8.1 &   3.4 &   9.7 &  1.7 &   75.4 &   33.4 & 0.8 & 0.4 &   15.2 &    6.9 \\
\hline
\hline
\end{tabular}
\begin{tablenotes}
   \item[a] Identification (ID) from \cite{Takekoshi_2017ApJ_835_55T}.
   \item[b] Equivalent linewidth from equation \ref{eq:equivalent_linewidth}.
   \item[c] The deconvolved radius ($R_{\text{deconv}}$) and total gas mass  ($M_{\text{gas}}$) from \cite{Takekoshi_2017ApJ_835_55T}. 
   We assume an uncertainty of $20\%$ in R.
   \item[d] $L_{\text{CO}}$ is in unit of $10^{2}$\,K\,\kms\,pc$^{2}$
   \item[e] Virial parameter $\alpha_{\text{vir}} = M_{\text{vir}}/M_{\text{gas}}$.
   \item[f] Dust-based $\alpha_{\text{CO}}$ in unit of \Xco.
\end{tablenotes}
\end{threeparttable}
\label{tab:SMC_dust_paramters}
\end{table*}


\subsubsection{Some Considerations and Caveats}

First, in the calculation of $\Sigma_{\rm gas}$,  \cite{Takekoshi_2017ApJ_835_55T} use GDR$=1000$ to determine the gas mass. \textcolor{black}{\bf \citet{RomanDuval_2014ApJ_797_86R} recommend GDR$\sim1800$ with a $\pm0.3$~dex uncertainty. \citet{Roman_Duval_2017ApJ_841_72R} find that the GDR is a function of surface density, with a factor of 7 change over $\Sigma_{\rm gas}\sim 10-100$\,M\sun\,pc$^{-2}$ in the direction of lower values in the denser ISM, although with significant methodological uncertainties. The GDR$=1000$ value is in good agreement with their Figure 14b for $\Sigma_{\rm gas}\sim100$\,M\sun\,pc$^{-2}$, but it is possible that the true GDR is larger, which would drive the dust clouds to be more bound. }

\textcolor{black}{Second, \citeauthor{Takekoshi_2017ApJ_835_55T} 
adopt a dust emissivity index $\beta = 1.2$ in the SED fit, and the dust emissivity can have a large impact on the dust-based total gas mass determination.} Commonly the dust emissivity follows a power law with the wavelength ($\kappa \propto \lambda^{-\beta}$), with power indexes ($\beta$) between $1$ and $2$. However, the dust emissivity is also a function of the dust grain properties, determined by coagulation, density, and temperature \citep{Ossenkopf_1994AA_291_943O,Bot_2010AA_524A52B,Paradis_2011AA_534A_118P}. The study done by \cite{Roman_Duval_2017ApJ_841_72R} analyzes the dust emissivity and GDR in the SMC, finding that matching depletion measurements may require reducing the GDR, although this could also be explained by biases in the UV absorption spectroscopy used to determine depletions introduced by the complex SMC velocity structure and also possibly Milky Way confusion. Adjustments in the direction of lower GDR would drive SMC dust clouds toward lower surface densities, increasing their virial parameter.

Third, another caveat in our determinations is the fact that there may be several \textcolor{black}{ CO clouds along the line of sight to a dust cloud, all of which will be included in our velocity dispersion calculation despite the fact that some may be chance alignments. This will artificially increase the velocity dispersion of the integrated spectrum associated with the dust cloud, and consequently its virial mass.} 
The inclusion of multiple velocity components would be correct if a dust cloud has a number of CO-emitting cores surrounded by a CO-faint envelope, but if some of these objects are not gravitationally associated we could overestimate the resulting virial mass. Correction for such an effect could drive some clouds to lower virial parameters.

\subsection{CO-to-H$_2$ conversion factor in the SMC}
\label{sec:conversion_factor}

Under the assumption of virial equilibrium, we can estimate the CO-to-H$_2$ conversion factor for the SMC through the ratio $\alpha^{\text{vir}}_{\text{CO}} = M_{\text{vir}}/L_{\text{CO}}$. In section \ref{sec:virial_mass_vs_Luminosity}, we find that this conversion factor is roughly constant with the luminosity for the SMC CO clouds (see Figure \ref{fig:Lco_scaling_relation}) identified by CPROPS. We determine a median value of the $\alpha^{\text{vir}}_{\text{CO}}$ of $\sim 10.5\,\pm\,5$ \Xco, at 9 parsec resolution, approximately $2.5$ times the canonical $\alpha_{\text{CO}}$ value in the inner disk of the Milky Way \citep[4.36\,\Xco,][]{Bolatto_2013ARA&A_51}.

This value of the  conversion factor for the SMC CO clouds is consistent with other virial mass-based conversion factors of $\approx 6-17$ \Xco\ calculated across the Magellanic Clouds at similar  resolution  \citep{Bolatto_2003ApJ_595_167B,Bolatto_2008ApJ_686,Herrera_2013AA_554A_91H,Muraoka_2017ApJ_844_98M}, and almost $\sim 2$ to $10$ times lower than those obtained  at coarser spatial resolution and  estimated using a similar method \citep{Rubio_1991_ApJ_368_173R,Rubio_1993AA_271_9R,Mizuno_2001PASJ_53L_45M,Muller_2010ApJ_712_1248M}. In fact,  \cite{Rubio_1993AA_271_9R}  observed that the conversion factor decreases with the cloud radius, approximately as $\alpha_{\text{CO}} \propto R^{0.7}$. As the median radius of the SMC clouds of this paper is $5.8$ pc (Table \ref{tab:median_parameters_smc}), a conversion factor of $\sim 13$  \Xco\ is predicted for the SMC by the \cite{Rubio_1993AA_271_9R} relation, which is consistent to our measurement. On smaller spatial scales we would expect the SMC $\alpha_{\text{CO}}$ decrease to smaller values similar to that of the Milky Way as we zoom into the CO-emitting core of a cloud. CO observations with spatial resolutions of $\sim$ 1 pc, which can be easily reached by the ALMA telescope, are required. However, recent ALMA studies of the SMC at such spatial scales \citep[$\sim 0.4-1.0$ pc,][]{Saldanio_2018_BAAA_60_192S,Valdivia_Mena_2020AA_641A_97V,Kalari_2020MNRAS_499_2534K} have shown that the virial mass-based conversion factor in the SMC and Magellanic Bridge is similar to the one derived in this study. This result suggests that $\alpha_{\text{CO}}$ in the SMC may have reached a constant value, independent of the spatial resolution and thus reflecting a conversion factor intrinsic to individual clumpy clouds.

We can estimate the CO-to-H$_{2}$ conversion factor using the gas mass determination from the dust emission of the CO clouds at 12 pc resolution by the equation \textcolor{black}{ $\alpha^{\text{dust}}_{\text{CO}} = M_{\text{H}_{2}}/L_{\text{CO}}$, where $M_{\text{H}_{2}} = M_{\text{gas}} - M_{\text{HI}}$}. 
For this estimate, we assume that the molecular component dominates the dust-based total gas mass over the atomic HI gas in the innermost part of the clouds. \textbf{This assumption \textcolor{black}{is consistent with PDR models of molecular clouds and supported} by the results by \cite{Jameson_2018_ApJ_853_111J} who find that the HI gas only contributes $\lesssim 5\%$ to the [CII] emission toward molecular regions in the SMC}. Thus, the dust-based conversion factor $\alpha^{\text{dust}}_{\text{CO}}$ {\bf (as an upper limit)} is roughly equal to $M_{\text{gas}}/L_{\text{CO}}$, where the gas mass is taken from \cite{Takekoshi_2017ApJ_835_55T}, and the CO luminosity is calculated by $L_{\text{CO}} = \pi\,R^{2}\,I_{\text{CO}}$. \textcolor{black}{Here,  $R$ is the radius of the dust emission and $I_{\text{CO}}$ is the CO intensity used in the equation \ref{eq:equivalent_linewidth} and measured in the CO APEX data convolved to 40" (12pc)  resolution.}
We obtain a median dust-based $\alpha^{\text{dust}}_{\text{CO}} = 28 \pm 15$ \Xco. This $\alpha^{\text{dust}}_{\text{CO}}$ is about $6.5$ times larger than the typical Milky Way disk conversion factor of $\alpha_{\text{CO}}=4.36$ \Xco\ \citep{Bolatto_2013ARA&A_51}.

\begin{figure}
	\includegraphics[width=\linewidth]{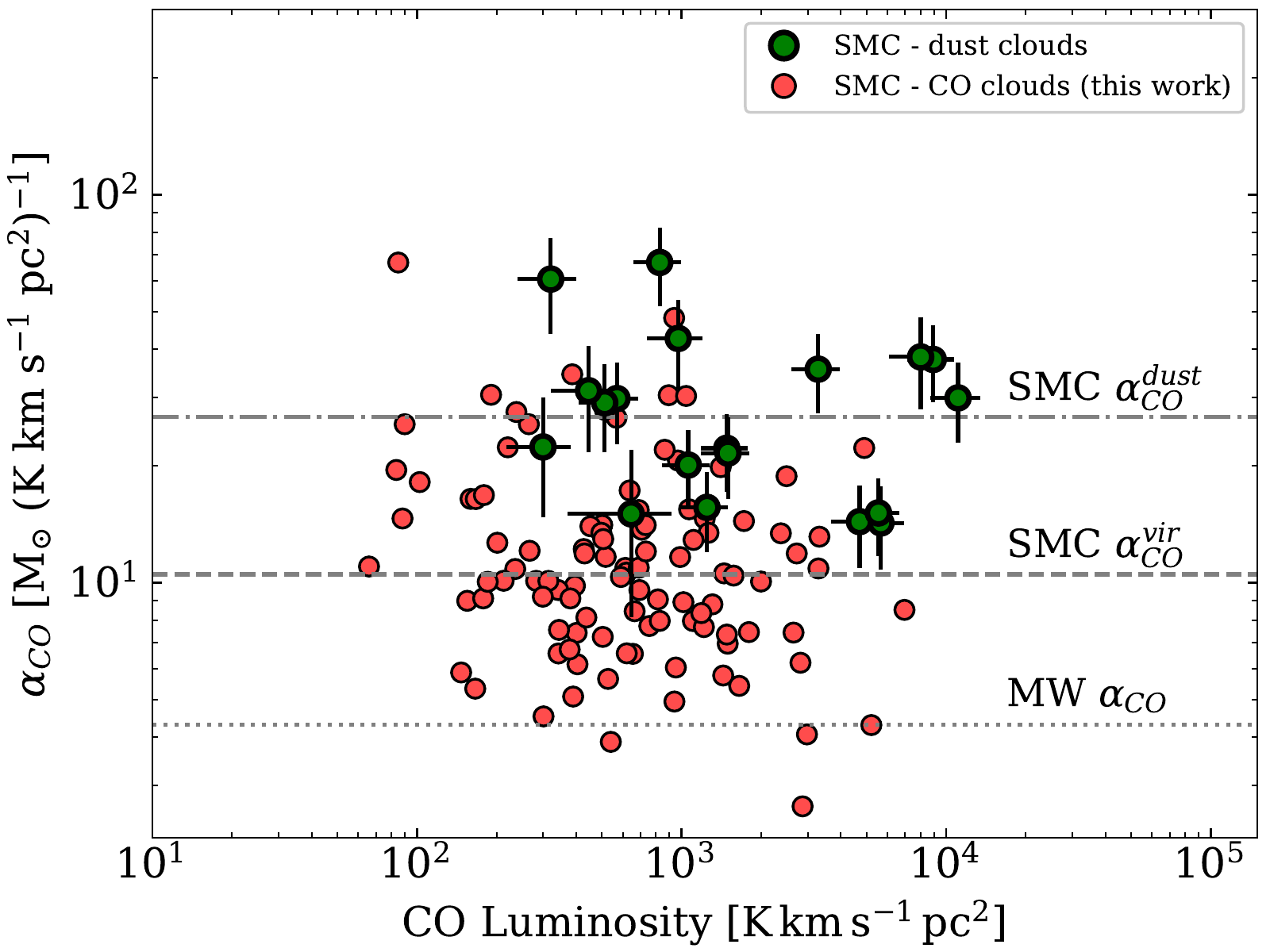}
    \caption{CO-to-H$_2$ conversion factor ($\alpha_{\text{CO}}$) as function of the CO luminosity for the 19 dust clouds (green dots) from \cite{Takekoshi_2017ApJ_835_55T} within our observed regions at 12 pc resolution. To calculate the CO luminosity, we use the measured CO intensity used in equation \ref{eq:equivalent_linewidth} and the dust cloud size ($R$) given by \cite{Takekoshi_2017ApJ_835_55T}. The dot-dashed line indicates the median dust-based $\alpha^{\text{dust}}_{\text{CO}} = 28\,\pm\,15$ \Xco\ for the SMC dust clouds. The conversion factor obtained for each of the 124 SMC CO clouds assuming gravitational virial condition are shown as red dots (see Figure \ref{fig:Lco_scaling_relation}d), with a median value of $\alpha^{\text{vir}}_{\text{CO}} = 10.5\,\pm\,5$ \Xco\ (dashed line). The  dotted line corresponds to the canonical Milky Way (MW) conversion factor.}
    \label{fig:dust_based_Xco}
\end{figure}

Figure \ref{fig:dust_based_Xco} shows the virial-based and the dust-based conversion factor as function of the CO luminosity. The dust-based conversion factor $\alpha^{\text{dust}}_{\text{CO}}$ estimated at 12 pc resolution for the SMC dust clouds lie above the median value derived from the virial method $\alpha^{\text{vir}}_{\text{CO}}$ (dashed line) for the 124 CO clouds at 9 pc resolution identified by CPROPS. There is no obvious trend of $\alpha^{\text{dust}}_{\text{CO}}$ or $\alpha^{\text{vir}}_{\text{CO}}$ with CO luminosity.

Note that since the virial parameter of our clouds is close to unity, $\alpha_{\rm vir}\sim1$ (Figure \ref{fig:Mvir-Mtot_relation}a), the virial mass on these scales is similar to the mass we obtain from dust. In other words, the $\alpha_{\rm CO}^{\rm dust}$ estimated on 12~pc scales is larger than the $\alpha_{\rm CO}^{\rm vir}$ on 9~pc scales. We will refer to the $\alpha_{\rm CO}$ on 12~pc scales as the ``dust'' value and that at 9~pc scales as the ``virial'' value, as the latter is only obtained through the virial estimate. This estimation of the dust-based conversion factor for the SMC clouds is consistent with previous calculations by using dust or [CII] emission in the SMC \citep{Leroy_2011ApJ_737_12L,Pineda_2017_ApJ_839_107P,Jameson_2016ApJ_825_12J,Jameson_2018_ApJ_853_111J}. 

The difference between these two values of $\alpha_{\rm CO}$ may also be due to uncertainties in the calculation, which are large. Alternatively, it is possible that the CO-emitting cores of molecular clouds are surrounded by a large envelope of CO-faint gas. In such a scenario the $\alpha_{\rm CO}$ measured at 9~pc reflects the mass-to-light ratio of the CO core and the material immediately surrounding it, while the measurement at 12~pc resolution is closer to the mass-to-light ratio of the core plus the envelope. Based on this interpretation, the total molecular gas mass of the core+envelope could be estimated using $M_{\text{mol}} = L_{\text{CO}}\,\alpha^{\text{dust}}_{\text{CO}}$, while the total core mass would correspond to $M_{\text{mol}} = L_{\text{CO}}\,\alpha^{\text{vir}}_{\text{CO}}$ using the total luminosity of $ L_{\text{CO}} \simeq 1.9\times10^{5}$ \Lco\ measured from cubes for SNR $> 3$. The masses computed using these methods are $5.3\times10^6$ \Msun\ and $2.0\times10^6$ \Msun\ respectively.

\subsection{Scaling Relation of Low Metallicity Galaxies}
\label{sec:scaling_relation_galaxies}

\begin{figure*}[ht!]
	\includegraphics{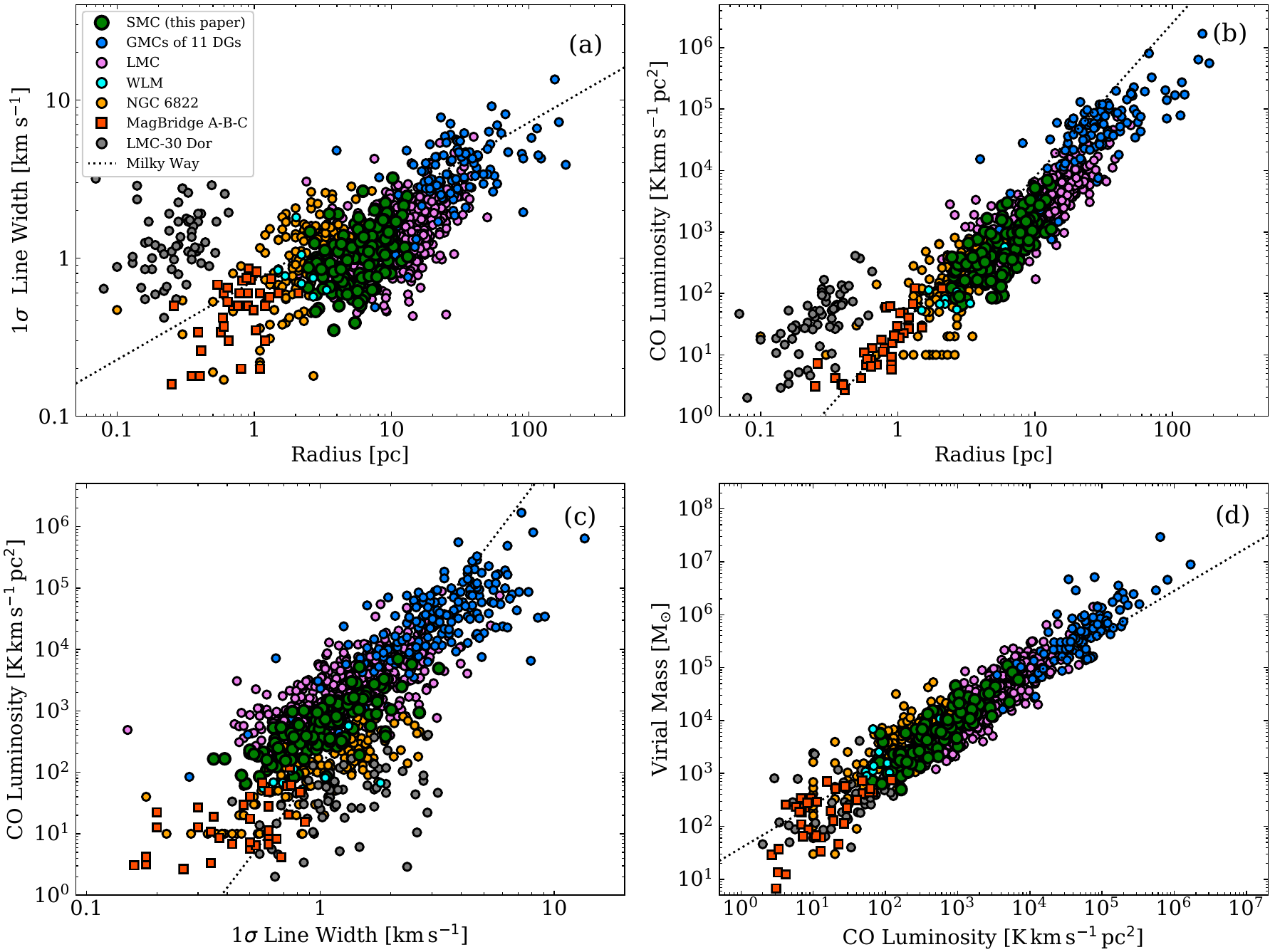}
    \caption{Scaling relation between the velocity dispersion, radius, luminosity, and virial mass of CO clouds from different low-metallicity galaxies. The SMC clouds (9 pc resolution), indicated in green dots, are compared with those clouds from Dwarf Galaxies (DGs) in blue dots \citep[$\sim 6-110$ pc resolutions,][]{Bolatto_2008ApJ_686}, the LMC clouds in pink \citep[11 pc resolution,][]{Wong_2011ApJS_197s}, WLM in cyan \citep[$\sim 5$ pc,][]{Rubio_2015_Nature_525_218R}, NGC 6822 in orange \citep[$\sim 2$ pc,][]{Schruba_2017ApJ_835_278S}, the Magellanic-Bridge A, B and C in red squares \citep[$\sim 1$ pc,][]{Saldanio_2018_BAAA_60_192S,Kalari_2020MNRAS_499_2534K,Valdivia_Mena_2020AA_641A_97V} and the LMC-30 Doradus in gray \citep[$\sim 0.45$ pc,][]{Indebetouw_2013ApJ_774_73I}. In this figure, we include all SMC clouds with SNR $> 3$. The dotted lines indicate the \cite{Solomon_1987ApJ_319_730S}'s relations for Milky Way GMCs. }
    \label{fig:scaling_relation_allGalaxies}
\end{figure*}

We compare the cloud properties that we measure for the SMC with those measured in several low metallicity galaxies such as the LMC at 11 pc resolution \citep{Wong_2011ApJS_197s} and 30 Doradus at 0.45~pc \citep{Indebetouw_2013ApJ_774_73I}, WLM \citep[$\sim 5$ pc resolution,][]{Rubio_2015_Nature_525_218R}, NGC 6822 \citep[$\sim 2$ pc resolution,][]{Schruba_2017ApJ_835_278S}, and the Magellanic Bridge regions \citep[$\sim 1$ pc,][]{Saldanio_2018_BAAA_60_192S,Kalari_2020MNRAS_499_2534K,Valdivia_Mena_2020AA_641A_97V}. These CO observations were done in the $J=1-0$ and $J=2-1$ transitions and analyzed using the CPROPS algorithm (with exception of WLM). The uniformity in the data sets, method, and analysis technique is very important to compare homogeneously molecular properties in different environments. We also use the \citet{Bolatto_2008ApJ_686}'s results, which span resolutions from $6$ to $110$ pc, and identify GMCs in different low metallicity galaxies with CPROPS. \textcolor{black}{The scaling relations between the velocity dispersion, radius, luminosity, and virial mass of CO clouds from different low-metallicity galaxies are shown in Figure \ref{fig:scaling_relation_allGalaxies}.} 

An overall feature is observed in the size-linewidth relation for the low metallicity galaxies: a departure to lower velocity dispersion (by a factor of $\sim 2$ for the SMC clouds) with respect to Milky Way clouds (dotted lines) of similar size, although clouds in NGC 6822 (yellow dots) appear to better correspond to the Galactic relation \citep[see][]{Schruba_2017ApJ_835_278S}. This departure could be a consequence of CO cloudlets not tracing the total turbulence of large H$_2$ envelopes as proposed by \cite{Bolatto_2008ApJ_686}. This is confirmed by the low value of CO-to-[CII] ratios \citep[$\sim$ 1/5 the fiducial value for the Milky Way,][]{Jameson_2018_ApJ_853_111J} found at low $A_{V}$, suggesting that there is a reservoir of H$_2$ gas mostly traced by [CII]\footnote{The [CII] 158 \mum\ emission is commonly used as the best tracer of "faint-CO" clouds at the edge of the molecular clouds since the lack of dust shielding will allow to the UV radiation field dissociate the CO, and most of the carbon will be as C$^{+}$.} emission \citep[$\sim$ 60-70\% of the molecular gas,][]{Requena_Torres_2016_AA_589A_28R,Jameson_2018_ApJ_853_111J} rather than CO. Alternatively, this shift in the velocity dispersion of SMC clouds could be due to a deficit of turbulent kinetic energy in the molecular clouds and thus other sources of energy (i.g., magnetic fields) would be required to support the CO clouds \citep[][]{Bot_2007AA_471_103B}. However, the gravitational virial equilibrium for the SMC clouds shown before in Section \ref{sec:stability_SMC}, indicates that there is no need for another physical component to keep the SMC clouds bound. On the other hand, \cite{Braine_2018_AA_612A_51B} noted that the low metallicity could also shift the size-linewidth relation to lower velocity dispersion in GMCs, but they do not distinguish whether the turbulent deficiency is due to changes in the metallicity or in the stellar surface density. This last scenario may be supported by numerical simulations \citep{Hocuk_2010_AA_510A_110H} which show that the fragmentation and the turbulence of a cloud ($4$ pc scale) can decay while the metallicity decrease.

At sub-parsec scales, the scatter in the linewidth increases considerably and may change significantly from region to region. In a quiescent region such as those in the Magellanic-Bridge \citep[red squares,][]{Saldanio_2018_BAAA_60_192S,Valdivia_Mena_2020AA_641A_97V,Kalari_2020MNRAS_499_2534K}, $\sigma_{\upsilon}$ is between $\sim 0.2-0.8$ \kms , while in the active 30 Doradus region \citep[gray dots,][]{Indebetouw_2013ApJ_774_73I} the velocity dispersion is between $\sim 0.8-5$ \kms. The large linewidths observed in 30 Doradus are suggestive of high stellar feedback and gravity-driven motions in high column density structures \citep[][]{Wong_2019ApJ_885_50W}. High linewidths with respect to Milky Way clouds for a given size have also been seen in active galaxies like the irregulars NGC 4526 and II Zw 40 \citep{Utomo_2015ApJ_803,Kepley_2016ApJ_828_50K} and the spirals M51, NGC 253, NGC 5253 \citep{Colombo_2014ApJ_784_3C,Leroy_2015_ApJ_801_25L,Miura_2018ApJ_864_120M}, all of them hosting starbursts.

In panels (b) and (c) of Figure  \ref{fig:scaling_relation_allGalaxies} corresponding to the luminosity-size and luminosity-linewidth relations, the offset of the SMC clouds from the Milky Way luminosity-scaling relations is also observed in low metallicity galaxies sample for radius larger than 1 pc (as discussed in Section \ref{sec:luminosity_scaling_relation}): these clouds tend to be less luminous for a given size and more luminous for a given linewidth. Conversely, for a fixed luminosity, the CO clouds of these irregular galaxies are larger in size and less turbulent than those of the Milky Way. Interestingly, for 30 Doradus the offsets in these luminosity scaling relations are in opposite directions, with clouds that are smaller but more turbulent than those predicted by \cite{Solomon_1987ApJ_319_730S} relation. Note, however, that the apparent offsets in both luminosity-scaling relations approximately cancel out in the virial mass calculation independently of the spatial resolution of the observations, so the correlation of CO-determined virial mass with CO luminosity (Panel d of Figure \ref{fig:scaling_relation_allGalaxies}) follows a trend very similar to that for the Milky Way. 

Despite the large scatter observed in Figure \ref{fig:scaling_relation_allGalaxies}, the scaling relations for low metallicity irregular galaxies appear to hold over a wide dynamic range between $\sim$ 1 to 120 pc, $\sim$ 0.5 to 10 \kms, $\sim 10^{1}$ to $10^{6}$ \Lco\ and $10^{2}$ to $10^{7}$ \Msun, although with some systematic departures with respect to the standard Milky Way scaling relations. 

\section{Summary and Conclusions}
\label{sec:summary}

We present a CO($2-1$) emission survey in the SMC with the  APEX single-dish telescope at $\sim9$~pc resolution. We identify a total of 177 clouds using the CPROPS algorithm \citep{Rosolowsky_2006PASP_118_590}, 124 of which are bright (SNRs $> 5$) and well-resolved clouds with radii between 2 to 13 pc, velocity dispersions of $0.4-3.2$ \kms, CO luminosities between $\sim 66-7\times10^{3}$ \Lco, and virial masses of $\sim 7\times10^{2}-10^{5}$ \Msun. \textbf{The  total luminosity of the identified 177 CO clouds  is $1.3\times10^{5}$ \Lco. Note that we estimate that  $\sim 30\%$ of the total CO luminosity in the investigated regions is not assigned to CO clouds.}

We find that, on average, the SMC clouds are less turbulent by a factor of $\sim 2$ than the inner Milky Way clouds \citep{Solomon_1987ApJ_319_730S} of similar size \citep[see also][]{Bolatto_2008ApJ_686}. Such departure is observed for all our SMC clouds, independently of the  region in which they are located. This feature is also shared with other low metallicity galaxies at scales larger than 1~pc. The most probable explanation is that the CO components are in the innermost regions of the entire molecular cloud \citep[][]{Bolatto_2008ApJ_686}. However, we also highlight that the poor metal abundance can be an important factor that contributes to the turbulent deficiency \citep{Braine_2018_AA_612A_51B}. 

We also find an agreement between the luminosity scaling relations of the SMC clouds with those of low-metallicity galaxies. Generally, clouds in low metallicity galaxies have larger sizes and smaller linewidths at a fixed CO luminosity than their inner Milky Way counterparts. On sub-parsec scales, the relations show increased scatter and higher departures, although all clouds tend to follow a fairly tight $L_{\text{CO}}-M_{\text{vir}}$ relation.

\textcolor{black}{We study the stability of the SMC clouds by comparing their virial masses (from our CO emission) with their dust mass from} 1.1~mm observations at 12 pc resolution \citep{Takekoshi_2017ApJ_835_55T}. We find that the SMC clouds are approximately in gravitational virial equilibrium, with the smaller clouds appearing less bounded. 

We estimate a virial-mass CO-to-H$_2$ conversion factor for the well-defined SMC clouds at 9 pc resolution. We obtain a median value of $\alpha^{\text{vir}}_{\text{CO(1-0)}} = 10.5\,\pm\,5$ \Xco. This conversion factor is 2.5 times the canonical value \citep[see][]{Bolatto_2013ARA&A_51}, \textcolor{black}{but we also show it is not inconsistent with the expectation for similarly luminous clouds in the Milky Way}. We estimate a dust-based CO-to-H$_2$ conversion factor for 19 dust clouds \citep{Takekoshi_2017ApJ_835_55T} within our observed area, finding a median of $\alpha^{\text{dust}}_{\text{CO(1-0)}} = 28\,\pm\,15$ \Xco. 

We find that the SMC clouds are mostly spatially associated with YSOs (38\%) and barely associated with HII regions (24\%). The YSOs associated with the CO clouds are luminous ($L_{*} \gtrsim 10^{3}$ \Lsun) and massive ($M_{*} \gtrsim 8$ \Msun), and most of them are in an early evolutionary stage (stage I) where accretion is considerable, while there is a poor association of clouds with more evolved YSOs (stages II/III). However, the HII regions would be interacting preferentially with large, luminous, and massive clouds. \textcolor{black}{\bf Our analysis also shows that bout half of the clouds are associated with ongoing star formation (either YSOs or HII regions), but also close to half the clouds are not.}

Finally, we obtain the cumulative mass distribution for the SMC clouds using both their luminous and virial masses. Fitting the mass distribution ($50$ clouds above the completeness limit), we find power-law exponent $\beta \lesssim -2$ in both mass distributions with a weak truncated end and a maximum mass of $M_0 = 3.8\times10^4$ \Msun\ for the luminous mass distribution, and essentially negligible truncation for the virial mass distribution with a $M_0 = 14.2\times10^4$ \Msun. The fitted maximum masses for the luminous and virial mass distributions of the SMC are about one order of magnitude lower than those of the LMC. Our results indicate that the molecular mass associated with CO clouds in the SMC is dominated by low-mass clouds across the galaxy.

\begin{acknowledgements}
      H.P.S acknowledges partial financial support from a fellowship from Consejo Nacional de Investigaci\'on Cient\'ificas y T\'ecnicas (CONCET-Argentina), and from Secretar\'ia de Ciencias y T\'ecnicas (SeCyT), C\'ordoba, Argentina, and partial support from ANID(CHILE) through FONDECYT grant No1190684. H.P.S. also thanks to E. Rosolowsky for sharing the IDL algorithm for the mass spectrum. M.R. wishes to acknowledge support from ANID(CHILE) through FONDECYT grant No1190684 and partial support from ANID project Basal AFB-170002 and Basal FB210003. A.D.B. acknowledges partial support from NSF-AST2108140.\\

This publication is based on data acquired with the Atacama Pathfinder Experiment (APEX) under program ID  (C093.F-9711A-2014) and (C.095F-9705A-2015) 
in Chile observing time. APEX is a collaboration between the Max-Planck-Institut fur Radioastronomie, the European Southern Observatory, and the Onsala Space Observatory.

\end{acknowledgements}

%
\bibliographystyle{aa} 
\bibliography{biblio2} 
%
\appendix

\section{CPROPS Aplication}
\label{app:cprops_aplication}

\textcolor{black}{We use the CPROPS to find the CO clouds (\cite{Rosolowsky_2006PASP_118_590}) within our data-cubes. For the decomposition method, we choose the modified CLUMPFIND algorithm (hereafter referred to as ECLUMP). As is explained by the authors,} the algorithm can be divided into two main parts, which are both signal identification and cloud decomposition. In the signal identification, the algorithm considers adjacent channels with intensities above a threshold (THRESH\,$\times\,\sigma_{rms}$) to make a mask, which then is expanded to include all emissions above an edge threshold (EDGE\,$\times\,\sigma_{rms}$) in the position-position-velocity cube (by default, the free parameters THRESH and EDGE have values of 4 and 2, respectively). In all our observed regions, we mostly used THRESH = 3 and EDGE = 1.7 for signal identification. \textcolor{black}{For a few clouds in the NE-Bar, SW-Bar, and DarkPK regions, an EDGE of 2.0 or 2.5 was used. On the contrary, for weak signals in three isolated regions in the NE-Bar, the EDGE parameter was reduced to a value of 1.2 or 1.5. In this step, we get masks with the most significant emissions, which are defined as "islands" and may be composed of local maxima (substructures) which have to be decomposed, and finally flagged as true or false clouds. We only consider the substructures as clouds those which have areas (MINAREA) larger than the beam size and velocity widths (MINVCHAN) $\ge$ 4 channels (1.0 \kms) in the decomposition.}

In some cases, we rescale the data to reduce the contrast between very close substructures to merge them into one cloud when the decomposition was not reliable (e.g., two very close unresolved clouds into one resolved cloud), preserving the relative brightness distribution. The parameter which handles the contrast is called BCLIP. We only used BCLIP = 2.4 for a few clouds in the SW-Bar. In DarkPK, the decomposition was forced to merge a couple of clouds making BCLIP = 0.2. 

CPROPS corrects for a substantial sensitivity bias by extrapolating each measured property of a cloud to find the value expected by perfect sensitivity (rms brightness temperature $T_{rms} = 0$ K). The extrapolation causes an enlargement effect on the parameter. Then, the measured properties are corrected for the resolution bias by using the beam size ($\sigma_{\text{beam}}$) and the width of the velocity channel through equations 9 and 10 from \cite{Rosolowsky_2006PASP_118_590}. It is important to note that both the cloud size and velocity dispersion are rather defined as rms sizes of the intensity distribution along the two spatial and spectral dimensions, respectively, and these values will increase as much higher are the data dispersion within the decomposed substructure.

\section{CO($2-1$) and Star-Forming Regions in the SMC}
\label{app:CO_yso_HII}

\begin{figure*}
    \centering
	\includegraphics[width=14cm]{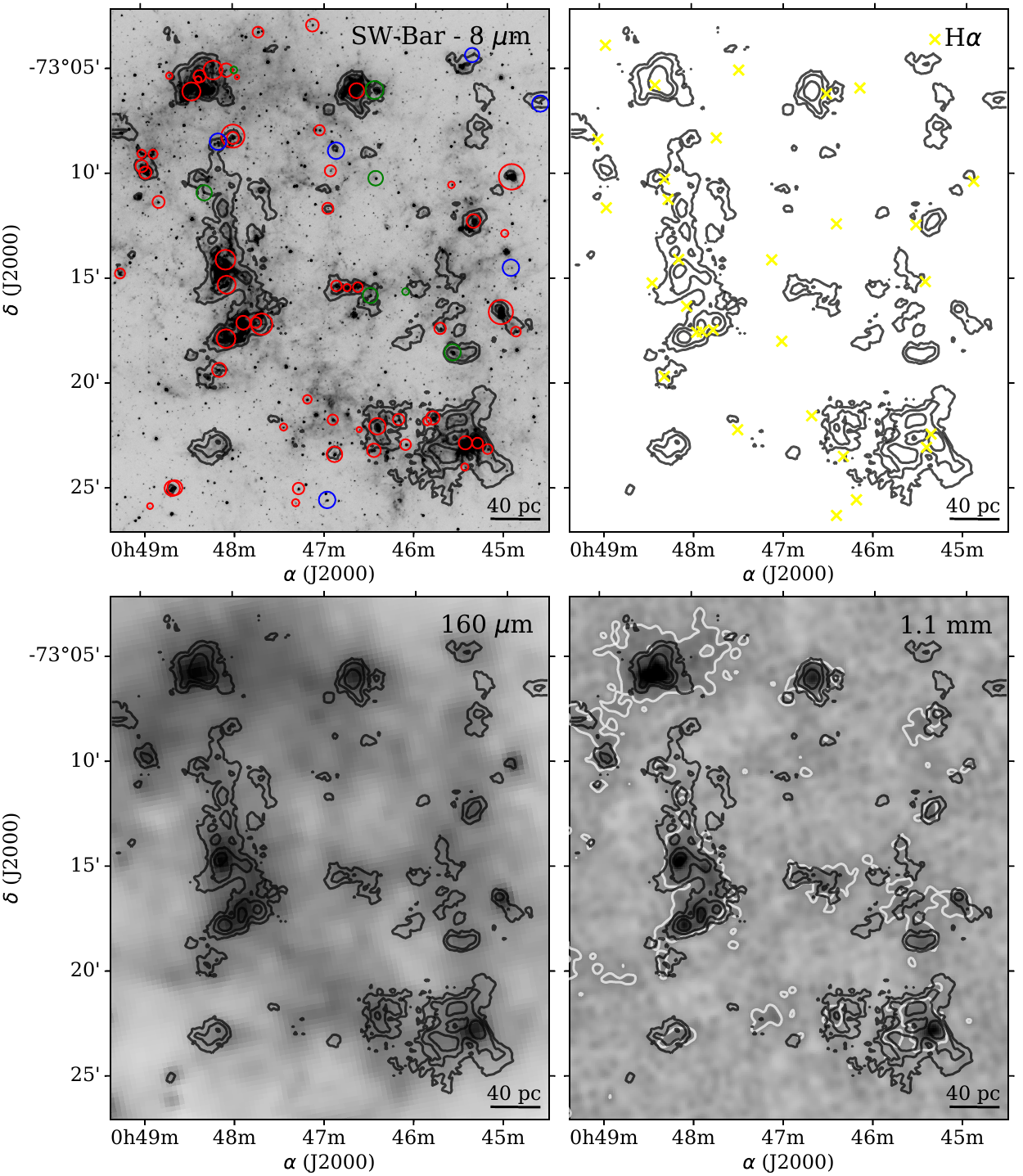}
    \caption{Superposition of the CO($2-1$) emission at 9 pc resolution (black contours) into the Spitzer 8 \mum\ map (upper left), MCELS H$\alpha$ map (upper right), Herschel 160 \mum\ map (bottom left), and AzSTEC 1.1 mm map (bottom right) towards the SW-Bar. The CO contours correspond to 1.5, 4.5, 10.0 K\,\kms. On the upper left panel, the YSOs Stage I, Stage II, and Stage III are indicated in red, green, and blue circles, respectively. The sizes of these circles are proportional to the stellar luminosity. On the upper right, the ``$\times$" symbols show the location of the HII regions. On the bottom right panel, the white contours indicate the 1.1 mm emission at 5$\sigma$.} 
    \label{fig:YSOs_dust_CO_emission_SW}
\end{figure*}

\begin{figure*}
    \centering
	\includegraphics[width=15.5cm]{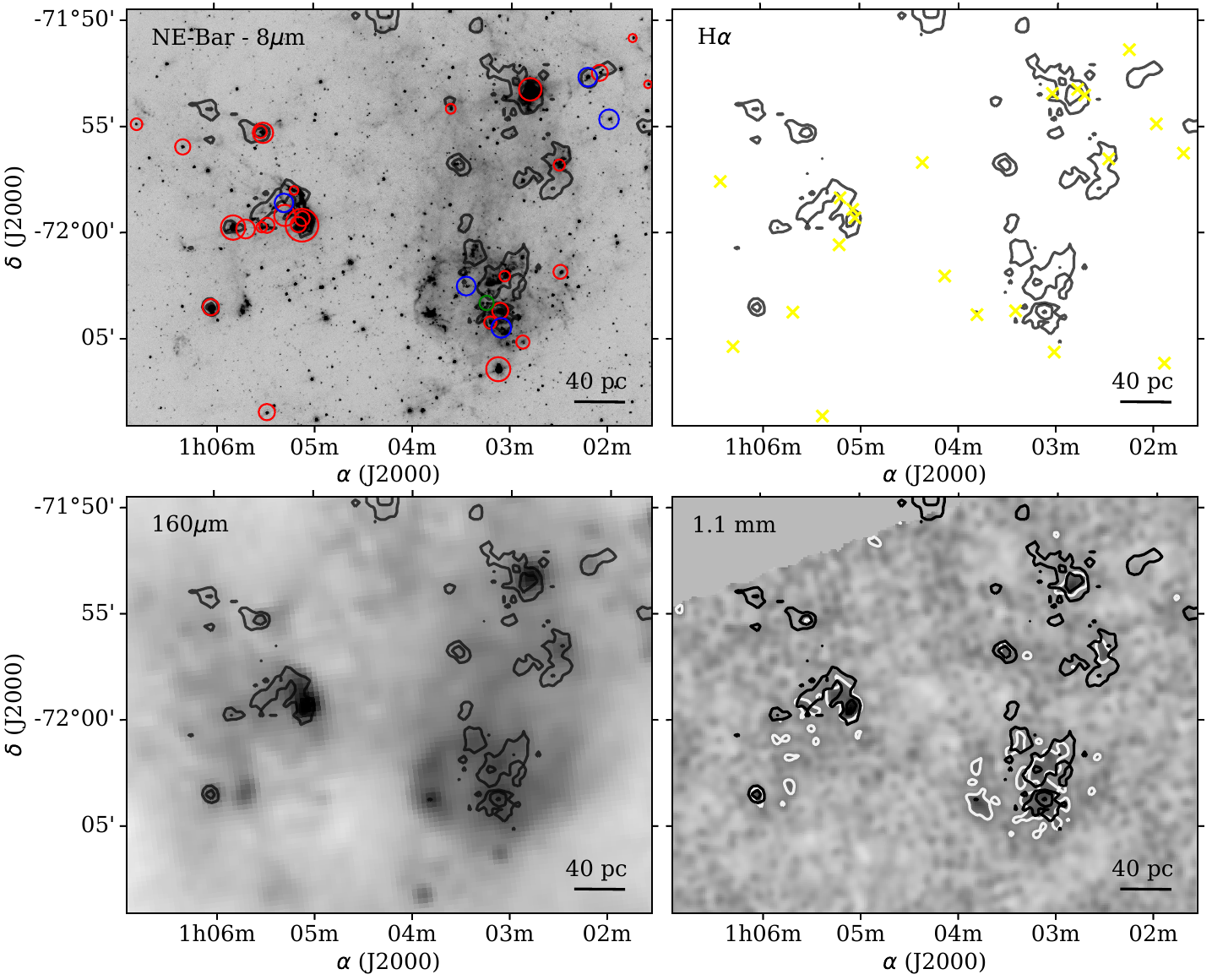}
    \caption{Similar to Figure \ref{fig:YSOs_dust_CO_emission_SW} but for NE-Bar, and contours of CO emission in 1.0, 2.0, 4.5 K\,\kms.}
    \label{fig:YSOs_dust_CO_emission_NE}
\end{figure*}

Figures \ref{fig:YSOs_dust_CO_emission_SW} to \ref{fig:dust_CO_emission_N88} show the good spatial correlation of the CO emission (black contours) with the star-forming regions traced by H$\alpha$, Spitzer 8 \mum, Herschel 160 \mum\ and 1.1 mm in the surveyed regions of the SMC. In general, the CO emission coincides spatially with peak emission at 8 \mum, 160 \mum, and 1.1 mm, indicating important star formation in these molecular clouds. The DarkPK region is an exception since weak or no emission is detected at these wavelengths (Figure \ref{fig:dust_CO_emission_DarkPK}), suggesting that the DarkPK is a quiescent region with low high-mass star formation \citep[see][]{Jameson_2018_ApJ_853_111J}. Similarly, some small CO clouds in the SW-Bar and NE-Bar have weak or no emission at any of these star-formation wavelength tracers, indicating quiescent molecular clouds. 

\textcolor{black}{We show the YSO candidates selected from \cite{Simon_2007ApJ_669_327S}, \cite{Carlson_2011ApJ_730_78C} and \cite{Sewilo_2013ApJ_778_15S} on  the 8 \mum\ maps of the observed regions.} These YSOs have been classified according to their evolutive sequential stages (I, II, and III) based on the modeling of their spectral energy distributions (SEDs) and the gas mass accretion criteria. We observe that the CO distribution agrees well with the youngest YSOs (Stage I) indicated with red circles, while a poorer association is found for more evolved YSOs (Stage II and III, green and blue circles, respectively). Comparing the position of the YSOs to the CO cloud center (see Section \ref{sec:CO_star-forming}), we find that 94 YSOs with defined evolutive stages are associated with CO clouds in our observed maps. Of these YSOs, $60$\% are Stage I YSOs, $16$\% are Stage II and III YSOs, and $24$\% are undefined YSOs. Although these fractions are dependent on the completeness of the existing YSOs catalogs as well as the sensitivity of CO observations, we expect that the Stage I YSOs keep showing high probabilities to be associated with molecular clouds. 

\subsection{Particular Regions of the SMC}
\label{app:CO_DUST_emission}

We describe the association of the star formation signposts with the molecular clouds in N66, N83, N88, and DarkPk regions observed in the SMC. 

\begin{figure*}
	\includegraphics[width=\linewidth]{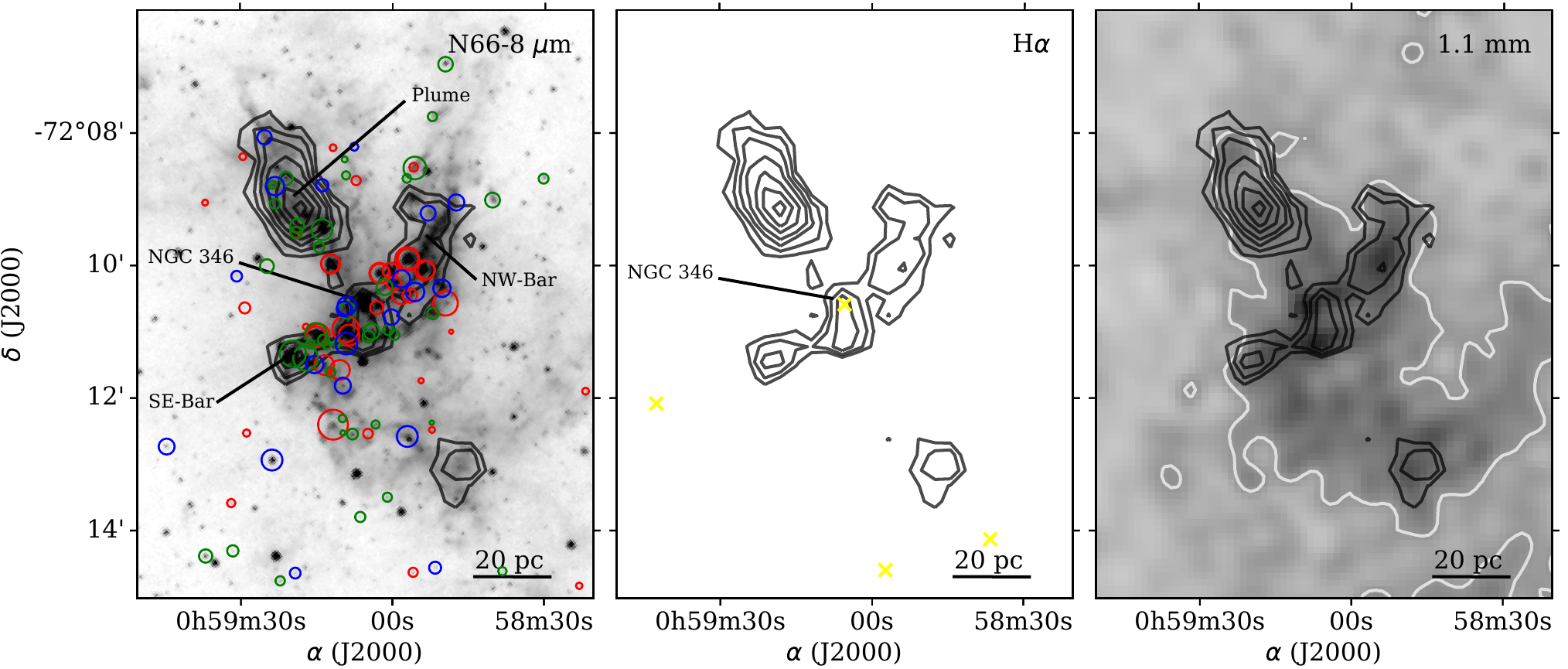}
    \caption{The CO distribution of N66 is shown in black contours, at levels 1.6, 2.6, 4.0, 5.7, 8.6, 11.4, and 14.3 K\,\kms, superimposed to the Spitzer 8 $\mu$m (left), MCELS H$\alpha$ (middle) and AzTEC 1.1 mm (right) maps. An evolutionary classification of YSOs from \cite{Sewilo_2013ApJ_778_15S} and \cite{Simon_2007ApJ_669_327S} are indicated by red (Stage I), green (Stages II) and blue (Stages III) circles. The sizes of these circles are proportional to the stellar luminosity. The main cluster of N66 (NGC 346), and the plume and bar structures are indicated. Yellow ``$\times$" symbols in the middle panel indicate the HII regions from \cite{Gouliermis_2012ApJ_748_64G}. In the right panel, the 1.1 mm emission at 5$\sigma$ is highlighted.}
    \label{fig:dust_CO_emission_N66}
\end{figure*}

{\bf N66:} This region is the brightest HII region in the SMC, characterized by extended gas emission ionized by the stellar cluster NGC 346. The CO gas distribution in the vicinity of NGC 346 follows an arc-like structure extending from the NW to the SE, associated with filaments at 8 \mum\ and peak emission at H$\alpha$ and 1.1 mm (see Figure \ref{fig:dust_CO_emission_N66}). In the same line of sight of NGC 346, we find two molecular clouds (No 105 and No 106 in Table \ref{tab:cprops_parameters_SMC}) despite the strong radiation field present. Both clouds, with high SNR $> 5$, show a large velocity separation of about 15 \kms\ of difference in their peak velocities and have the widest linewidths ($\sigma_{\upsilon} \approx 2$ \kms) of the NE clouds (see Figure \ref{fig:COparam_histo}). These large linewidths in both clouds may indicate a tight interaction with NGC 346, whereas the large separation in velocity may confirm that the ISM in N66 seems to be lacerated by ionized gas as suggested by \cite{Hony_2015MNRAS_448_1847H}. \textcolor{black}{Embedded young massive stars have been spectroscopically identified in the NIR towards CO clouds in N66 \citep{Rubio_2018AA_615A_121R} }

Perpendicular to the arc-like bar in the NE direction, there is the strongest, densest but cold CO gas (plume-like structure) in N66 \citep[see][]{Rubio_2000AA_359_1139R,Requena_Torres_2016_AA_589A_28R} with strong emission at 8 \mum\ (not associated with Stage I YSOs), but weak at H$\alpha$ and 1.1 mm. Finally, a weak CO cloud is detected to the SW of N66, likely a quiescent cloud since no association with YSOs is found, and only weak emissions at 8 \mum\ and 1.1 mm are detected.

\begin{figure*}
	\includegraphics[width=\linewidth]{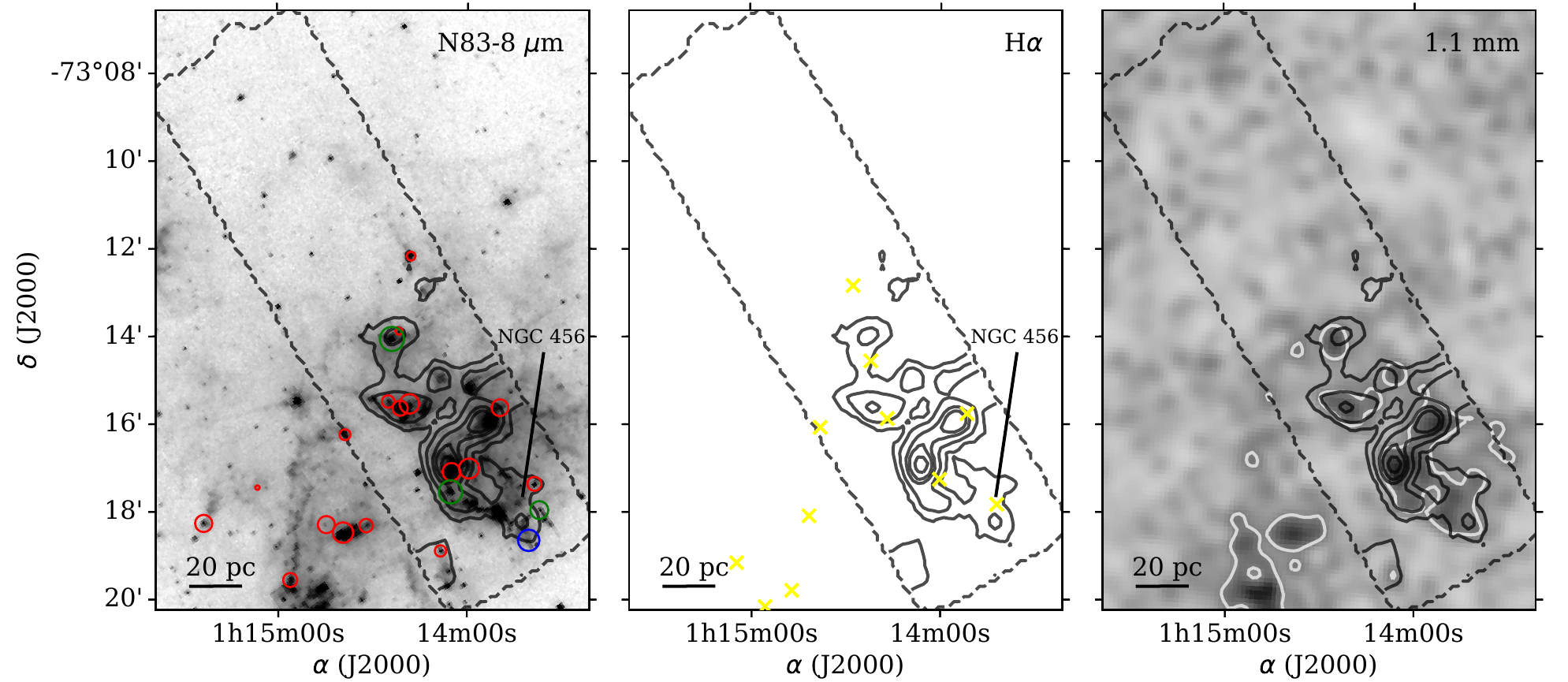}
    \caption{The CO distribution of N83 is shown in black contours, at levels 1, 3, 7, 10, and 15 K\,\kms, superimposed to the Spitzer 8 $\mu$m (left), MCELS H$\alpha$ (middle), and AzTEC 1.1 mm (right) maps. The evolutionary classification of YSOs from \cite{Sewilo_2013ApJ_778_15S} is indicated by red (Stage I), green (Stages II), and blue (Stages III) circles. The sizes of these circles are proportional to the stellar luminosity. Yellow ``$\times$" symbols in the middle panel indicate the HII regions from \cite{Gouliermis_2012ApJ_748_64G}. In the right panel, the 1.1 mm emission at 5$\sigma$ is highlighted.}
    \label{fig:dust_CO_emission_N83}
\end{figure*}

{\bf N83:} This is an isolated, relatively active star formation region with several molecular clouds cataloged. Some of them lie between two extended H$\alpha$ emissions, one of them hosts the NGC 456 cluster in the southernmost part of N83 (Figure \ref{fig:dust_CO_emission_N83}). There are other four more compact HII regions in the area (indicated with yellow "$\times$" symbols) that are closer to the brightest CO clouds. Two of these clouds (No 158 and No 160 on Table \ref{tab:cprops_parameters_SMC}) have the highest CO luminosity of the WG clouds, with $L_{\text{CO}} > 2\times10^{3}$ \Lco). These clouds could be associated with an expanding shell of NGC 456, although there is no clear evidence whether these bright clouds are being swept up by the expanding shell of the HII region NGC 456 or they are only ambient clouds interacting with the ionized gas \citep{Bolatto_2003ApJ_595_167B}. The third most luminous cloud, with $L_{\text{CO}} \simeq 1.2 \times10^{3}$ \Lco\ (cloud No 165), would be associated with the HII region located towards the North-East direction of NGC 456.

In particular, these very luminous clouds show smaller radius and  linewidth than the SMC clouds of similar luminosity, and in consequence lower $M_{\text{vir}}$ than their comparable clouds in luminosity (see Figure \ref{fig:Lco_scaling_relation}). Such features may be attributed to their interaction with the expanding shell of the HII regions in N83.

\begin{figure*}
	\includegraphics[width=\linewidth]{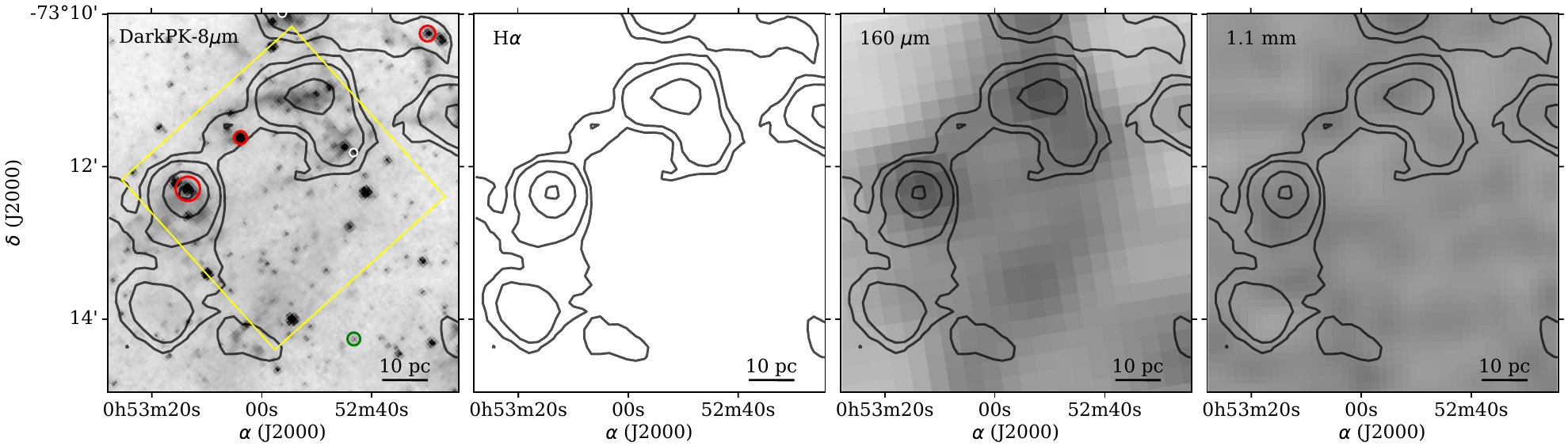}
    \caption{The CO distribution of DarkPK is shown in black contours, at levels 0.5, 1.0, 3.0, and 6.0 K\,\kms, superimposed to the Spitzer 8 $\mu$m, MCELS H$\alpha$, Herschel 160 $\mu$m and AzTEC 1.1 mm maps. The evolutionary classification of YSOs from \cite{Sewilo_2013ApJ_778_15S} is indicated by red (Stage I) and green (Stages II) circles. The sizes of these circles are proportional to the stellar luminosity. The yellow box in the 8 \mum\ map indicates the regions where CO emission was detected with ALMA by \cite{Jameson_2018_ApJ_853_111J}.} 
    \label{fig:dust_CO_emission_DarkPK}
\end{figure*}

{\bf DarkPK:} This region is one of the more quiescent regions of the SMC. Figure \ref{fig:dust_CO_emission_DarkPK} shows the CO distribution slightly correlated with the 8 \mum\ emission, and not correlated with H$\alpha$ emission, while only one YSO (Stage I) is associated with a CO clump. At longer wavelengths, the CO clouds show a better association with warm dust at 160 \mum\ rather than cool dust emission not detected at 1.1 mm \citep{Takekoshi_2017ApJ_835_55T}. These associations correlate well with the finding from \cite{Jameson_2018_ApJ_853_111J} in the DarkPK. These authors observed low levels of CO($2-1$) and atomic ([CII], [OI]) emissions. They derive
an averaged $N_{\text{H}2} \approx 0.5\times10^{21}$ \cmc, which is $\sim$ 2 to 6 times lower than those of other CO-bright regions into the SW-Bar. They also estimate a very low $A_{\text{V}}$ with a mean value of $\sim$ 0.8 mag, which is consistent with the very low CO and H$_2$ abundance. 

\begin{figure*}
	\includegraphics[width=\linewidth]{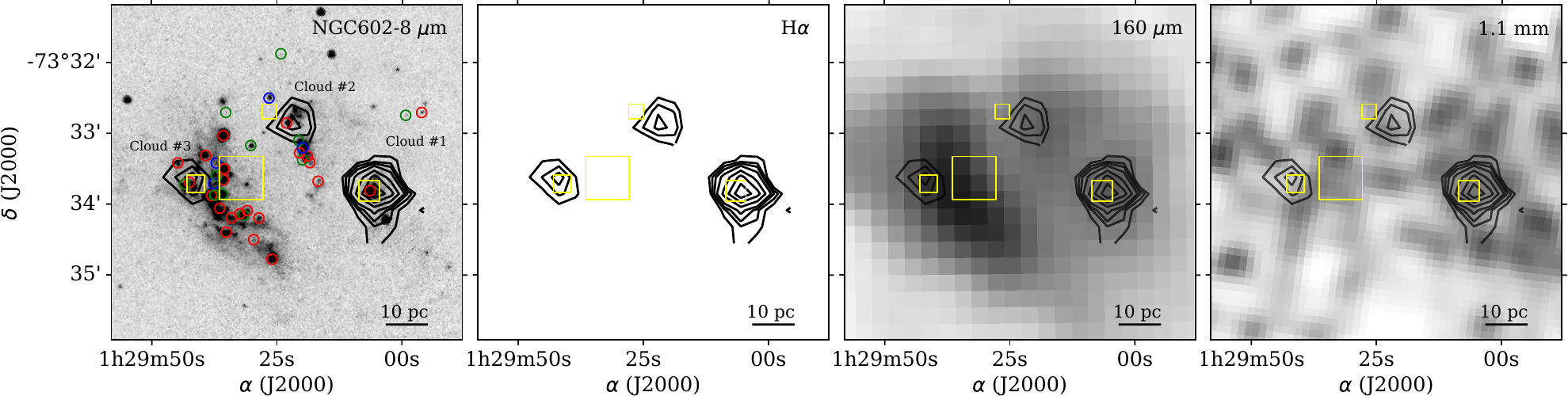}
    \caption{The CO distribution of NGC 602 is shown in black contours, at levels 0.4, 0.5, 0.6, 0.9, 1.3, 1.7, and 2.1 K\,\kms, superimposed to the Spitzer 8 $\mu$m, MCELS H$\alpha$, Herschel 160 $\mu$m and AzTEC 1.1 mm maps. The evolutionary classification of YSOs from \cite{Carlson_2011ApJ_730_78C} are indicated by red (Stage I), green (Stages I/II and II), and blue (Stages II/III and III) circles. The largest yellow square represents the main stellar cluster and the smaller ones PMS sub-clusters \citep[see][]{Gouliermis_2012ApJ_748_64G} associated with the CO emission.}
    \label{fig:dust_CO_emission_NGC602}
\end{figure*}

{\bf NGC 602/N90:} this is an HII complex characterized by two ridges of dust filaments towards the South-East and North-West (see Figure \ref{fig:dust_CO_emission_NGC602}), and an ionized cavity by the bright young stellar cluster NGC 602 (big yellow box). The region harbors tens of massive YSOs \citep[see][]{Carlson_2011ApJ_730_78C}, most of them of Stage I over the ridges. Our three CO clouds in the outskirt of the main cluster coincide with the location of Stage I YSOs and smaller pre-main-sequence (PMS) sub-clusters \citep[small yellow boxes, see][]{Gouliermis_2012ApJ_748_64G}. The central cluster NGC 602 seems to have blown all the molecular gas beyond an average distance of $\sim$ 12-16 pc \citep[see][]{Carlson_2011ApJ_730_78C}, where the YSOs and CO gas are located \citep[see also][]{Fukui_2020arXiv200513750F}. The association of the three CO clouds with warm dust emission detected at 160 \mum\ (denser in the SE ridge) may explain the shielding of our molecular component. The high radiation field produced in the vicinity of NGC 602 may be heating the environment at considerable temperatures so that cold dust emission at 1.1 mm is not detected in the region. In fact, cold dust emissions are extremely rare in the Wing and Magellanic Bridge as was shown by \cite{Takekoshi_2017ApJ_835_55T} and \cite{Valdivia_Mena_2020AA_641A_97V}.

\begin{figure*}
	\includegraphics[width=\linewidth]{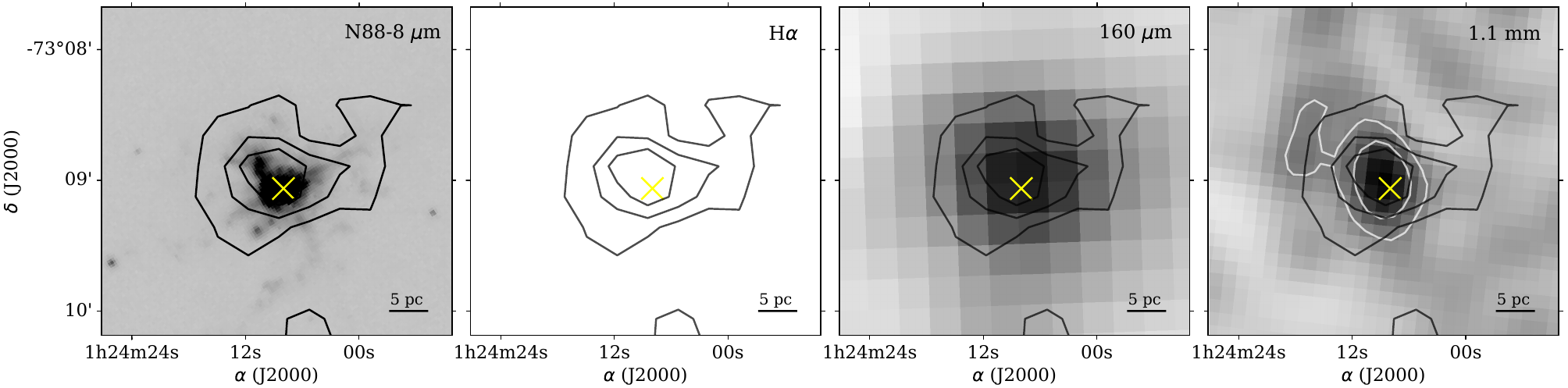}
    \caption{The CO distribution of N88 (A) is shown in black contours, at levels 1.4, 2.8, 4.3 K\,\kms, superimposed to the Spitzer 8 \mum\ band (top left), MCELS H$\alpha$ (top right), Herschel 160 \mum\ (bottom left), and AzTEC 1.1 mm (bottom right) maps. While in white contours, the continuum 1.1 mm emission is highlighted. The ``$\times$" symbols indicate the position of the ultra-compact HII region N88-A.}
    \label{fig:dust_CO_emission_N88}
\end{figure*}

{\bf N88:} In this region, only one molecular cloud was found. This cloud is in the line-of-sight of a strong source at 8 \mum\ and  H$\alpha$ (Figure \ref{fig:dust_CO_emission_N88}). Most of the short-wavelength emission comes from the ultra-compact (UC) HII region N88-A (indicated with an "$\times$" symbol) excited by zero-age main-sequence (ZAMS) massive stars of spectral-type O6 \citep{Ward_2017MNRAS_464_1512W}. Spectroscopic studies in Br$\gamma\,\, 2.1661$ \mum, and H$_{2}\,\, 2.1218$ \mum\ emission-lines \citep{Ward_2017MNRAS_464_1512W} show an arc-like H$_2$ structure that is associated with an expanding ionized Br$\gamma$ gas whose line emissions are peaking at heliocentric centroid velocities of $153-156$ \kms\ ($V_{lsr} = 144-147$ \kms). The CO cloud in this region has a peak emission at $V_{lsr} = 147.6$ \kms\ (cloud No 174 in Table \ref{tab:cprops_parameters_SMC}), quite similar to that of the main Br$\gamma$ component (\#35A) at $V_{lsr} = 146.8$ \kms\  \citep[][]{Ward_2017MNRAS_464_1512W}. Therefore, it is most likely that the UC-HII region and the CO cloud are associated. As expected, the molecular cloud is associated with compact emission at 160 \mum\ and 1.1 mm, consistent with  CO being protected from the high radiation field of the UC compact ionized region. In fact, the high H$_2$ density in this region \citep[$\sim 3-10$ times higher than that in N66, see][]{Requena_Torres_2016_AA_589A_28R} estimated from [CII] emission, would prevent the fast photodissociation of the CO cloud in the most extreme environments of the SMC.

\end{document}